\documentclass[a4paper,fleqn,usenatbib]{mnras}
\usepackage[T1]{fontenc}
\usepackage{ae,aecompl}
\usepackage{graphicx}
\usepackage{amsmath}
\usepackage{amssymb}
\usepackage{txfonts}
\usepackage{slashed}
\usepackage{makecell}

\newcommand{\specialcell}[2][c]{%
  \begin{tabular}[#1]{@{}l@{}}#2\end{tabular}}
  
\usepackage{caption}
\usepackage{multicol}

\usepackage[dvipsnames]{xcolor}
\definecolor{armygreen}{rgb}{0.0, 0.6, 0.4}

\hypersetup{pdfauthor={A. Spurio Mancini \textit{et el.}},
               pdftitle={Testing (modified) gravity with 3D and tomographic cosmic shear},
              pdfkeywords={gravitational lensing: weak -- dark energy},
               bookmarksnumbered=true}

\usepackage{etoolbox}
\makeatletter
\patchcmd\@combinedblfloats{\box\@outputbox}{\unvbox\@outputbox}{}{
  \errmessage{\noexpand\@combinedblfloats could not be patched}
}
\makeatother

\newcommand{\ltsima}{$\; \buildrel < \over \sim \;$}
\newcommand{\lsim}{\lower.5ex\hbox{\ltsima}}
\newcommand{\gtsima}{$\; \buildrel > \over \sim \;$}
\newcommand{\gsim}{\lower.5ex\hbox{\gtsima}}

\newcommand{\dd}{\mathrm{d}}

\usepackage[section]{placeins}

\title[Testing (modified) gravity with 3D and tomographic cosmic shear]{Testing (modified) gravity with 3D and tomographic cosmic shear}

\author[A. Spurio Mancini et al.]{
A. Spurio Mancini$^{1}$\thanks{Contact e-mail: \href{mailto:spuriomancini@thphys.uni-heidelberg.de}{spuriomancini@thphys.uni-heidelberg.de}},
R. Reischke,$^{2}$
V. Pettorino,$^{3,4}$
B.M. Sch\"afer$^{2}$
and M. Zumalac\'{a}rregui$^{5,6}$
\\
$^{1}$Institut f\"ur Theoretische Physik, Heidelberg Universit\"at, Philosophenweg 12, 69120 Heidelberg, Germany\\
$^{2}$Astronomisches Rechen-Institut, Zentrum f{\"u}r Astronomie der Universit{\"a}t Heidelberg, Philosophenweg 12, 69120 Heidelberg, Germany\\
$^{3}$Astrophysics Department, IRFU, CEA, Universit\'{e} Paris-Saclay, F-91191, Gif-sur-Yvette, France\\
$^{4}$Universit\'{e} Paris-Diderot, AIM, Sorbonne Paris Cit\'{e}, CEA, CNRS, F-91191 Gif-sur-Yvette, France\\
$^{5}$Berkeley Center for Cosmological Physics, LBNL and University of California at Berkeley, Berkeley, California 94720, USA\\
$^{6}$Institut de Physique Th\'{e}orique, CEA, Universit\'{e} Paris-Saclay, CNRS, F-91191 Gif-sur-Yvette, France}
\date{}
\begin{document}
\pagerange{\pageref{firstpage}--\pageref{lastpage}}
\pubyear{2017}
\maketitle
\label{firstpage}

\setlength{\parindent}{10pt}

\begin{abstract}
Cosmic shear is one of the primary probes to test gravity with current and future surveys. There are two main techniques to analyse a cosmic shear survey; a tomographic method, where correlations between the lensing signal in different redshift bins are used to recover redshift information, and a 3D approach, where the full redshift information is carried through the entire analysis. Here we compare the two methods, by forecasting cosmological constraints for future surveys like Euclid. We extend the 3D formalism for the first time to theories beyond the standard model, belonging to the Horndeski class. This includes the majority of universally coupled extensions to $\Lambda$CDM with one scalar degree of freedom in addition to the metric, still in agreement with current observations. Given a fixed background, the evolution of linear perturbations in Horndeski gravity is described by a set of four functions of time only. We model their time evolution assuming proportionality to the dark energy density fraction and place Fisher matrix constraints on the proportionality coefficients. We find that a 3D analysis can constrain Horndeski theories better than a tomographic one, in particular with a decrease in the errors of the order of 20$\%$. This paper shows for the first time a quantitative comparison on an equal footing between Fisher matrix forecasts for both a fully 3D and a tomographic analysis of cosmic shear surveys. The increased sensitivity of the 3D formalism comes from its ability to retain information on the source redshifts along the entire analysis. 
\end{abstract}

\begin{keywords}
weak lensing -- dark energy -- modified gravity
\end{keywords}

\section{Introduction}

The observed acceleration of the Universe \citep{Riess1998, Perlmutter1999} can be ascribed to a Dark Energy component accounting for approximately $70\%$ of the energy budget of the Universe. From a theoretical point of view, identifying the Dark Energy with a cosmological constant term $\Lambda$ fits well the observations, but has been questioned in terms of naturalness and interpretation in terms of energy density of the vacuum \citep[see][for a recent review]{Martin2012}. Alternatives to the cosmological constant can be generally grouped into two main categories. Either Dark Energy is a modification of gravity on the largest scales (``modified gravity'' theories), or it is given by a scalar field that effectively behaves as a fluid with negative pressure (usually referred to as proper ``dark energy'' models). The distinction between these two classes can at times be feeble \citep[see][for a recent discussion]{Joyce2016} and the vast amount of proposed theories \citep[see][for a review]{Clifton2012} urgently calls for methods to be developed, aiming at distinguishing among the large number of theoretical options with advanced statistical methods and efficient computational effort. This is particularly relevant in light of the unprecedented amount of data that will come from many space- and ground-based experiments, such as Euclid \footnote{https://www.euclid-ec.org/}, SKA \footnote{https://www.skatelescope.org/}, LSST \footnote{https://www.lsst.org/} and WFIRST \footnote{https://wfirst.gsfc.nasa.gov/}, whose launch in the next few years is planned with the goal of unveiling the true nature of the cosmic acceleration.

The Horndeski class of modified gravity theories represents an example of a remarkably large set of extensions to General Relativity. First discussed in 1974 by Horndeski \citep{Horndeski1974} and subsequently rediscovered in \citet{Nicolis2009} and \citet{Deffayet2011}, the Horndeski Lagrange density is the most general gravitational theory with one scalar degree of freedom, in addition to the metric tensor, with derivatives in the equations of motion not higher than second order; this guarantees safety from ghost-like degrees of freedom. This set of theories collects under its name many different models of dark energy/modified gravity
(see Sec.~\ref{sec:Horndeski} for a list of some of them). Theories that contain higher order derivatives, but are still free from ghost degrees of freedom, belong to the `Beyond Horndeski' category \citep{Zumalacarregui2014, Gleyzes2015, Langlois2017, Crisostomi2018}.

On the observational side, many different probes have been proposed to investigate dark energy/modified gravity models. These include type Ia Supernovae, Baryon Acoustic Oscillations, galaxy clustering and weak gravitational lensing, to name a few \citep[see e.g.][for an exhaustive review of the different probes]{Weinberg2013}. In this paper we focus on the weak gravitational lensing caused by the large-scale structure of the Universe, or cosmic shear. Since the first detections in early 2000s \citep[e.g.][]{Bacon2000, VanWaerbeke2000, Brown2003}, this field has developed within a well-established theoretical and experimental framework. Cosmic shear is particularly appealing as one of the most promising probes of dark energy \citep{Jain2003, Bernstein2004, Hannestad2006, Amendola2008, Huterer2010}: the differential deflection in light bundles from distant galaxies caused by variations of the gravitational fields of the large-scale structure result in a coherent distortion of galaxy images as we observe them on the sky \citep[see][for reviews on the topic]{Bartelmann2001, Hoekstra2008, Kilbinger2013}. Thus cosmic shear is sensitive to the growth rate of the perturbations of the gravitational potential and to the geometry of the Universe through the distance-redshift relation. These features are crucial for dark energy studies, as they allow us to measure how the expansion rate and growth of structure change with time: it follows that the sensitivity of cosmic shear to dark energy can be fully exploited if the analysis performed is able to recover information on the evolution in redshift of the large-scale structure. This is only to a little extent achieved in a 2-dimensional analysis: galaxy shapes are observed on the 2-dimensional celestial sphere and the shear components are line-of-sight projected quantities, with the projection causing loss of information on the redshift evolution \citep{Jain1997, Takada2003b, Takada2003a, Munshi2006, Jee2013, Kilbinger2013}. For this reason, as an alternative to a pure 2D projection, a tomographic analysis based on a binning in redshift of the sources has been first proposed in \citet{Hu1999} and has since become the standard technique for cosmological weak lensing studies \citep{Takada2004a, Simon2004, Takada2004b, Hollenstein2009, Kilbinger2009, Schaefer2012, Heymans2013}. Galaxies are assigned to different bins according to their redshifts, so that intra- and inter-bin correlations of the binned shear field can be computed. This reduces the range of the projection to the width of the bins and allows for some gain in redshift information through the inter-bin correlations. Despite its success in providing some sensitivity to the growth of structure with its `2D$\frac{1}{2}$' nature, as it has sometimes been relabelled, tomography has still the disadvantage of representing a compression of data: the 2D analysis performed within a single bin is such that the range of the projection is smaller than in the pure 2D case, being restricted to the width of the bin, however this does not represent yet a fully 3D treatment of the shear field. 

As an alternative to tomography, a method to retain information on the redshift of each source galaxy along the entire weak lensing analysis has been first proposed in \citet{Heavens2003} and subsequently refined in \citet{Castro2005}, \citet{Heavens2006} and \citet{Kitching2011}. Based on a spherical Fourier-Bessel decomposition of the shear field, it is immune from the aforementioned approximations presented in \citep{Kitching2016a}. 
In addition to avoiding any binning and averaging in redshift, the spherical Fourier-Bessel formalism allows for a separation of angular ($\ell$) and radial ($k$) modes \citep{Kitching2014}: the fact that they can be treated independently makes it easier than in tomography to reduce the impact of problematic small scales, where models for the non-linear growth of structure \citep{Smith2003, Takahashi2012, Mead2015} or baryon feedback \citep{semboloni2011, vanDaalen2011, Semboloni2013} do not yet provide a fully reliable description. These advantages compensate for the extra computational time required by the more complicated integrations in the covariance of the shear modes. 
To date, the 3D weak lensing approach has been applied to real data only in \citet{Kitching2007, Kitching2014, Kitching2016b} for a $\Lambda$CDM model. \citet{Grassi2014} investigated the possibility of detecting baryon acoustic oscillation features in the cosmic matter distribution by 3D weak lensing; \citet{Zieser2016} studied the cross-correlation between the 3D weak lensing signal and the integrated Sachs-Wolfe effect; \citet{Camera2011} investigated the constraining power of 3D cosmic shear on a class of Unified Dark Matter models, where a single scalar field mimics both dark matter and dark energy, whereas \citet{Ayaita2012} employed 3D cosmic shear to explore the capability of future surveys to constrain dark energy clustering. While 3D weak lensing has been partially studied in the context of modified gravity theories in \citet{Pratten2016} to constrain $f(R)$ chameleon models and environmentally dependent dilaton models, showing that for an all-sky spectroscopic survey the $f(R)$ parameter $f_{R_0}$ can be constrained in the range $f_{R_0} < 5 \times 10^{-6} (9 \times 10^{-6})$ for $n = 1(2)$ with a 3$\sigma$ confidence level, there has not been any application to a larger class of modified gravity theories. \citet{Alonso2016} forecast the sensitivity of future surveys to Horndeski theories using different probes, among which tomographic weak lensing.

In this paper we propose for the first time 3D cosmic shear as a probe of Horndeski theories of modified gravity. We analyse in detail the expected performance of a Euclid-like experiment, with the aim of forecasting the precision with which future stage IV surveys will be able to constrain this class of alternatives to General Relativity using cosmic shear datasets. We choose the parametrization of linear perturbations in Horndeski gravity first proposed by \citet{BelliniSawicki2014} and based on four functions of time only, which completely describe the evolution of linear perturbations once the background evolution is fixed. We model their time evolution assuming proportionality to the density fraction of dark energy and constrain the proportionality coefficients under the assumption of Gaussian likelihood. We simultaneously also place constraints on a set of standard cosmological parameters describing the evolution of the background, including the sum of the neutrino masses.

We produce our forecasts for both a fully 3D and a tomographic analysis of the measurements, with the aim of comparing the performances of the two methods on both modified gravity and standard cosmological parameters. \citet{Kitching2011} showed the relationship between weak lensing tomography and the 3D cosmic shear field, connected by the Limber approximation, a harmonic-space transform and a discretization in wavenumber. Our work presents for the first time a quantitative comparison on an equal footing between 3D and tomographic techniques for cosmic shear in terms of Fisher forecasts, showing that the 3D approach has more sensitivity than tomography to both standard and modified gravity cosmological parameters. 
We vary both the background cosmological parameters and those describing Horndeski theories, and consider only weak gravitational lensing as a cosmological observable, to test its power in constraining modified gravity theories without other probes and compare 3D and tomographic methodologies. 

This article is structured as follows: in section \ref{sec:Horndeski} we review the Horndeski Lagrangian and introduce the \citet{BelliniSawicki2014} parameterization; in section \ref{sec:cosmicshear} we present both the spherical Fourier-Bessel formalism for 3D weak lensing and the tomographic approach, in a modified gravity context; in section \ref{sec:Fisher} we explain our Fisher matrix analysis and report the specifications of the Euclid-like survey we consider; in section \ref{sec:selected_results} we present our Fisher forecasts for both 3D and tomographic cosmic shear; in section \ref{sec:conclusions} we draw conclusions from our analysis.

\section{Horndeski theories}\label{sec:Horndeski}

The Horndeski Lagrange density \citep{Horndeski1974} is the most general way of writing the Lagrangian of a scalar-tensor theory of gravity that is four-dimensional, Lorentz-invariant, local, and has equations of motion with derivatives not higher than second order. This ensures the safety of the theory against Ostrogradski instabilities and subsequent ghost degrees of freedom \citep{Woodard2007}. The Horndeski action can be written as follows:

\begin{align}\label{eq:Horndeski_action}
S [ g_{\mu \nu}, \phi ] &= \int \mathrm{d}^4 x \sqrt{- g} \left[ \sum_{i=2}^5 \frac{1}{8 \mathrm{\pi} G_N} \mathcal{L}_i[g_{\mu \nu}, \phi] + \mathcal{L}_m[g_{\mu \nu}, \psi_M] \right],\\
\mathcal{L}_2 &= G_2 (\phi, X), \nonumber \\
\mathcal{L}_3 &= -G_3(\phi, X) \Box \phi, \nonumber \\
\mathcal{L}_4 &=  G_4 (\phi, X) R + G_{4X}(\phi, X) \left[ (\Box \phi)^2 - \phi_{;\mu\nu} \phi^{;\mu\nu} \right],\nonumber \\
\mathcal{L}_5 &= G_5 (\phi, X) G_{\mu\nu} \phi^{;\mu\nu} \nonumber \\ 
&- \frac{1}{6}G_{5X} (\phi, X) \left[ (\Box \phi)^3 + 2 \phi_{;\mu}{}^{\nu} \phi_{;\nu}{}^{\alpha} \phi_{;\alpha}{}^{\mu} - 3 \phi_{;\mu\nu} \phi^{;\mu\nu} \Box \phi \right]. \nonumber
\end{align}  
The four contributions $\mathcal{L}_i$ of the gravitational sector depend on arbitrary functions of the metric $g_{\mu \nu}$ and the kinetic term $K = -\frac{1}{2} \partial _\mu \phi \partial^{\mu} \phi$ of the additional scalar degree of freedom $\phi$. The subscripts $\phi, X$ denote partial derivatives, e.g. $G_{iX} = \frac{\partial G_i}{\partial X}$. We write the normalization of the $G_i$ functions following the convention implemented in the \texttt{hi\char`_class} code \citep{Zumalacarregui2016}. We will consider only universal coupling between the metric and the matter fields (collectively described by $\psi_m$ and contained in the matter Lagrangian $\mathcal{L}_m$), which are therefore uncoupled to the scalar field. Most of the universally coupled models with one scalar degree of freedom belong to the Horndeski class. These include for example quintessence \citep{Ratra1988, Wetterich2009}, Brans-Dicke models \citep{Brans1961}, $k$-essence \citep{Armendariz1999, Armendariz2001}, kinetic gravity braiding \citep{Deffayet2010, Kobayashi2010, Pujolas2011}, covariant galileons \citep{Nicolis2009, Deffayet2009}, disformal and Dirac-Born-Infeld gravity \citep{deRham2010a, Zumalacarregui2013, Bettoni2013}, Chameleons \citep{Khoury2004b, Khoury2004a}, symmetrons \citep{Hinterbichler2010, Hinterbichler2011}, Gauss-Bonnet couplings \citep{Ezquiaga2016} and models screening the cosmological constant \citep{Charmousis2012, MartinMoruno2015}. Archetypal modified gravity-models such as all variants of $f(R)$ \citep{Carroll2004} and $f(G)$ \citep{Carroll2005} theories are also included.
Models that are not within this broad class are those that contain higher derivatives in the equations of motion \citep{Zumalacarregui2014, Gleyzes2015}, and modifications of gravity with non-scalar degrees of freedom, e.g. Einstein-Aether models \citep{Jacobson2001} or ghost-free massive gravity \citep{deRham2010b, deRham2011, Hassan2012}. The choice of the $G_i(g_{\mu \nu}, K)$ functions completely specifies the single modified gravity model that one considers. 

When dealing with linear perturbations acting on a Friedmann-Robertson-Walker metric in modified gravity one can assume spatial flatness and, considering only scalar perturbations \citep[see][for vector and tensor perturbations]{Durrer2016, Adamek2016}, write the line element in Newtonian gauge as
\begin{align}
\label{def:pot}
\mathrm{d}s^2 = -\left(1+2\frac{\Phi}{c^2} \right) c^2 \mathrm{d}t^2 + a^2 \left( t \right) \left(1-2\frac{\Psi}{c^2} \right) \mathrm{d} \mathbf{x}^2
\end{align}
with the Bardeen potentials $\Phi$ and $\Psi$. In General Relativity $\Phi = \Psi$ in absence of anisotropic stress, while this is in general not true in modified gravity. In \citet{Gleyzes2013} and \citet{BelliniSawicki2014} it has been shown that one can parametrize the evolution of linear cosmological perturbations in Horndeski theories by means of four functions of (conformal) time only, which we will collectively refer to here as $\alpha$ functions. Each of them carries a physical meaning, which we describe briefly here, referring to \citet{BelliniSawicki2014} and references therein for a more complete description:
\begin{itemize}
\item $\alpha_K$ is the \textit{kineticity} term, i.e. the kinetic energy of the scalar perturbations arising directly from the action. Increasing this term suppresses the sound speed of scalar perturbations. This makes the sound horizon smaller than the cosmological horizon, allowing the scalar field to enter a quasi-static configuration on smaller scales, below the sound horizon \citep{Sawicki2015}. In the quasi-static approximation where time derivatives are considered to be sub-dominant with respect to space derivatives, $\alpha_K$ does not enter the equations of motion and is therefore largely unconstrained by cosmic shear (\citealp{Bellini2016, Alonso2016}, although see~\citealp{Kreisch2017}).

\item $\alpha_B$ is the \textit{braiding} term, which describes mixing of the scalar field with the metric kinetic term, leading to what is typically interpreted as a fifth force between massive particles.

\item $\alpha_M$ is the \textit{Planck-mass run rate}, defined by 
\begin{align}
\alpha_M \equiv \frac{\mathrm{d \, ln}M_*^2}{\mathrm{d \, ln} \, a},
\end{align}
where $M_*^2$ is the dimensionless product of the normalization of the kinetic term for gravitons and $8 \pi G_N$ measured on Earth. This function describes the rate of evolution of the effective Planck mass.

\item $\alpha_T$ is the \textit{tensor speed excess,} indicating deviations from the speed of light in the propagation speed of gravitational waves. This can lead to anisotropic stress even in the absence of scalar field perturbations, as a result of a change in the response of the Newtonian potential to matter sources. Recently, very strong constraints have been placed on $\alpha_T$ by the measurement of the gravitational waves speed derived by the detection of the binary neutron star merger GW170817 and the gamma ray burst GRB170817A \citep{Abbott2017b, Abbott2017a, Baker2017, Creminelli2017, Ezquiaga2017, Sakstein2017, Lombriser2017, Bettoni2017}. Since the speed has been found to be very close to the speed of light, $\alpha_T$ has been consequently constrained to be very close to zero at the present time. We remark that the other three functions (as well as $\alpha_T$'s past value), are instead still free to vary. \cite{Ezquiaga2017} identify the models within the Horndeski classes that are still viable after GW170817; \cite{Peirone2018} show that, even with the strict bound on the present-day gravitational wave speed, there is still room within Horndeski theories for nontrivial signatures of modified gravity that can be measured at the level of linear perturbations. 
\end{itemize}

The specific model considered within the Horndeski class is defined by the choice of the $\alpha_i$ functions. The $\Lambda$CDM model corresponds to the choice $\alpha_K = \alpha_B = \alpha_M = \alpha_T = 0.$
Once the $\alpha$ functions are set, \citet{BelliniSawicki2014} show that it is sufficient to solve the equations of motion for the background and perturbations to fully determine the evolution of linear perturbations at the linear level.

In this paper, we aim at forecasting constraints on the $\alpha$ functions with a Fisher matrix analysis for a cosmic shear survey. In doing so, we need to choose a parametrization for the time evolution of the $\alpha$ functions. Following a common procedure, already implemented in \texttt{hi\char`_class}, we choose to parametrize these functions such that they trace the evolution of the dark energy component, to which they are proportional
\begin{align}\label{eq:alpha_time_evolution}
\alpha_i = \hat{\alpha}_i \, \Omega_{\mathrm{DE}}(\tau).
\end{align}
This choice, first suggested in \citet{BelliniSawicki2014}, is the simplest and the most common in the literature \citep[as used e.g. in][]{PlanckDE2016} and, despite not being the only one, can already provide a lot of information on Horndeski gravity, as remarked by \citep{Gleyzes2017} who showed that simple parametrizations are sufficient to describe the theory space in Effective Field Theory of Dark Energy \citep{Gubitosi2013, Gleyzes2015}, which the \citet{BelliniSawicki2014} parameterization belongs to. The forecasts we present in Sec.\ref{sec:selected_results} will therefore be on the proportionality coefficients $\hat{\alpha}_i$.

\section{Cosmic shear}\label{sec:cosmicshear}
\subsection{3D cosmic shear}\label{sec:3Dcosmicshear}
Weak gravitational lensing of the large-scale structure in the Universe is the deflection of light rays coming from distant sources due to the distortion of the cross-sectional shape of light bundles \citep[see][for reviews on the topic]{Bartelmann2001, Hoekstra2008, Kilbinger2015}. The typical lensing signal is a distortion, or shear, of the source image due to the gravitational potential generated by the mass distribution between the source and the observer. Cosmic shear measurements are of statistical nature, as the lensing effect is not associated with a particular intervening lens, but rather corresponds to small distortions (of the order of 1$\%$) by all density fluctuations along the line of sight. The statistical properties of the shear reflect those of the underlying density field and a particularly informative insight is given in configuration space by the two-point statistics or correlation function of the shear field. 

We present here a general formalism for a fully 3D expansion of the shear field that does not perform any binning in redshift. This is based on a spherical Fourier-Bessel decomposition of the shear, first introduced in lensing studies by \citet{Heavens2003}. Here we follow the notation and conventions of \citet{Zieser2016} and extend the presentation given there to a general modified gravity scenario characterised by the Bardeen potentials $\Phi$ and $\Psi$ defined in eq.\ref{def:pot}. 

Information on the gravitational potential is encoded in a weighted projection along the line of sight, the lensing potential. In a modified gravity context, considering perturbations at the linear level, the lensing potential $\phi$ is related to the Bardeen potentials $\Psi$ and $\Phi$ by
\begin{align}\label{eq:lensing_potential}
\phi (\chi, \hat{\mathbf{n}}) = \int_0^{\chi} \mathrm{d}\chi' \frac{\chi-\chi'}{\chi \chi'} \frac{\Phi(\chi, \hat{\mathbf{n}})+\Psi(\chi, \hat{\mathbf{n}})}{c^2}, 
\end{align} 
where $\chi$ is a comoving distance, and the normalized vector $\hat{\mathbf{n}}$ selects a direction on the sky. Here and throughout the paper spatial flatness will be assumed, and the integration in \eqref{eq:lensing_potential} is carried out in Born approximation, i.e. along the unperturbed light path. The shear tensor $\gamma (\chi, \hat{\mathbf{n}})$ is defined as the second $\slashed{\partial}$-derivative \citep{Newman1962, Goldberg1967} of the lensing potential \citep{Heavens2003, Castro2005}
\begin{align}\label{eq:edth-derivative}
\gamma (\chi, \hat{\mathbf{n}}) = \frac{1}{2} \slashed{\partial} \slashed{\partial} \phi (\chi, \hat{\mathbf{n}}).
\end{align}
The $\slashed{\partial}$-derivative acts as a covariant differentiation operator on the celestial sphere and relates quantities of different spin, raising the spin $s$ of a function, a number which characterises its transformation properties under rotations. Acting twice on $\phi$, the $\slashed{\partial}$ operator relates the scalar (spin-0) lensing potential to the spin-2 shear field $\gamma$.
The shear $\gamma$ can be expanded with a choice of basis functions given by a combination of spherical Bessel functions $j_\ell(z)$ \citep{Abramovitz1988} and spin 2-weighted spherical harmonics $_2 Y_{\ell m}(\hat{\mathbf{n}})$
\begin{align}\label{eq:sphericalFourier-Bessel}
\gamma (\chi, \hat{\mathbf{n}}) = \sqrt{\frac{2}{\pi}} \sum_{\ell m} \int k^2 \, \mathrm{d} k \, \gamma_{\ell m}(k) \, _2 Y_{\ell m} (\hat{\mathbf{n}}) \, j_\ell(k \chi),
\end{align} 
where the coefficients $\gamma_{\ell m}(k)$ are given by
\begin{align}\label{eq:coefficients_gamma}
\gamma_{\ell m}(k) = \sqrt{\frac{2}{\pi}} \int \chi^2 \mathrm{d}\chi \int \mathrm{d}\Omega \, \gamma(\chi, \hat{\mathbf{n}}) \, j_\ell(k \chi) \, _2 Y_{\ell m}^*(\hat{\mathbf{n}}).
\end{align}
Inserting \eqref{eq:lensing_potential} and \eqref{eq:edth-derivative} in \eqref{eq:coefficients_gamma}, and applying a spherical Fourier-Bessel expansion to the Bardeen potentials $\Phi$ and $\Psi$, we can rewrite $\gamma$ as
\begin{align}
&\gamma(\chi, \hat{\mathbf{n}}) = \sqrt{\frac{2}{\pi}} \frac{1}{c^2} \int_0^{\chi} \mathrm{d}\chi' \, \frac{\chi-\chi'}{\chi \chi'} \\
&\times \int k^2 \mathrm{d}k  \sum_{\ell m} \sqrt{\frac{(\ell+2)!}{(\ell-2)!}} \left[\frac{\Phi_{\ell m}(k, \chi')+\Psi_{\ell m}(k, \chi')}{2}\right] j_\ell(k \chi') \, _2Y_{\ell m}(\hat{\mathbf{n}}),\nonumber
\end{align}
where the division by 2 comes from the prefactor in \eqref{eq:edth-derivative}.
 
Poisson's equation can be used to link the coefficients in the spherical Fourier-Bessel decomposition of the lensing potential to those of the overdensity field $\delta_{\ell m}(k, \chi)$,
\begin{align}\label{eq:poisson}
\frac{\Phi_{\ell m}(k,\chi)}{c^2} = - \frac{3}{2} \frac{\Omega_m}{(k \chi_H)^2}\frac{\delta_{\ell m}(k,\chi)}{a(\chi)} \mu(k, a(\chi)),
\end{align}
with the Hubble radius $\chi_H \equiv c/H_0$. Here the function $\mu(k, a(\chi))$ describes the mapping from the potential fluctuations to the density fluctuations. Eq. (\ref{eq:poisson}) can also be used as a parametrization of modified gravity theories \citep[e.g.][]{PlanckDE2016}. The latter approach, however, only holds in the quasi-static regime, where one neglects terms involving time derivatives in the Einstein equations for perturbations and keeps only spatial derivatives \citep{Sawicki2015, Baker2015}. For the Euclid survey, it could be questionable if this approximation holds, given the large scales in principle accessible by the survey. The validity of the quasi-static approximation depends also on the single modified gravity model considered and its predictions for the sound speed of the additional scalar degree of freedom. That said, we stress that we are not using this parametrization, but rather take the potential and density statistics directly from \texttt{hi\char`_class}, which does not use the quasi-static approximation.

The density field is statistically homogeneous and isotropic, characterised by a power spectrum which is diagonal in harmonic space
\begin{align}
\left\langle \delta_{lm}(k, z) \delta_{\ell'm'}^{*}(k', z') \right\rangle = \frac{P_{\delta}(k, z, z')}{k^2}\delta^D(k-k') \delta_{\ell \ell'}^K \delta_{mm'}^K.
\end{align}

Using this, we can relate the covariance of shear modes to the matter power spectrum by
\begin{align}\label{eq:3Dcovariance}
\left\langle\bar{\gamma}_{lm}(k)\bar{\gamma}_{\ell'm'}^*(k')\right\rangle &= \frac{9 \Omega_m^2}{16 \pi^4 \chi_H^4}\frac{(\ell+2)!}{(\ell-2)!}\\
&\times \int \frac{\mathrm{d}\tilde{k}}{\tilde{k}^2} \, G_\ell(k, \tilde{k}) \, G_\ell(k', \tilde{k}) \, \delta_{\ell \ell'}^K \, \delta_{m m'}^K.\nonumber
\end{align} 
where
\begin{align}
G_\ell(k,k') &= \int \mathrm{d}z \, n_z(z) \, F_\ell(z,k) \, U_\ell(z,k'),\label{eq:G}\\ 
F_\ell(z,k) &= \int \mathrm{d}z_p \, p(z_p|z) \, j_\ell[k \chi^0(z_p)],\label{eq:F}\\ 
U_\ell(z,k) &= \frac{1}{2}\int_0^{\chi(z)} \frac{\mathrm{d} \chi'}{a(\chi')} \frac{\chi-\chi'}{\chi \chi'} j_\ell(k \chi') \, P_\delta^{1/2} \left( k, z \left( \chi \right) \right)\label{eq:U}\\ 
&\times \mu(k, a(\chi)) \left[1 + \frac{1}{\eta(k, a(\chi'))}\right] \nonumber.
\end{align}
$\bar{\gamma}$ are estimates of the shear modes that, in addition to the pure lensing effect, keep into account the redshift distribution of galaxies and the redshift estimation error (see sec.\ref{subsec:observational} for details on observational effects), as evidenced by the definition of the quantities in \ref{eq:G},\ref{eq:F},\ref{eq:U}, which contain the redshift distribution $n_z(z)$ of the lensed galaxies and the conditional probability $p(z_p|z)$ of estimating the redshift $z_p$ given the true redshift $z$. 
\begin{table*}
\begin{center}
 \begin{tabular}{cccccccc}
  \hline
  $z_m$ & $n_0[\mathrm{arcmin}^{-2}]$ & $\sigma_z$ & $\Omega_\mathrm{survey}[\mathrm{deg}^2]$ & $\ell_{\mathrm{min}}$ & $\ell_{\mathrm{max}}$ & $k_{\mathrm{max}}$ & $n_\mathrm{bins}$ \\ 
  \hline\hline
  0.9 & 30 & 0.05 & 15000 & 10 & 1000 & 1.0 & 10 \\
  \hline
 \end{tabular}
\end{center}
\caption{Specifications used in the Fisher matrix analysis for the Euclid survey: the median redshift $z_m$; the source density $n_0$; the error in photometric redshifts, $\sigma(z) = \sigma_z (1+z)$; the field size $\Omega_{sky}$. $\ell_{\mathrm{min}}$, $\ell_{\mathrm{max}}$ and $k_{\mathrm{max}}$ describe instead the minimum and maximum radial modes and the maximum angular mode, respectively, considered in the computation of the shear covariances \eqref{eq:3Dcovariance} and the Fisher matrix \eqref{eq:Fishermatrix}. $n_\mathrm{bins}$ is the number of bins considered in the tomographic analysis.}
 \label{tab:specifications}
\end{table*}
These two elements and the lensing kernel contained in the function $U_\ell(z,k)$ introduce correlations between the amplitudes of the signal on different scales; the covariance matrix then acquires off-diagonal terms, the calculation of which is numerically involved. The basis of spherical Bessel functions leads to integrals with rapidly oscillatory kernels, which have to be solved for a large number of parameter combinations.
$\eta(k, a(\chi'))$ is defined as the ratio between the Bardeen potentials
\begin{align}\label{eq:eta}
\eta(k, a(\chi)) = \frac{{\Phi(k, a(\chi))}}{{\Psi(k, a(\chi))}}.
\end{align}
The $P_\delta^{1/2} \left( k, z\left( \chi \right) \right)$ term comes from an approximation, introduced and justified in \citet{Castro2005}, to calculate unequal-time correlators appearing in the comoving distance integrations by means of a geometric mean $P \left( k, z, z' \right) \simeq \sqrt{P \left( k, z \right) P \left( k, z' \right)}$  \citep[see also][]{Kitching2017}. This expression simplifies considerably in the linear regime of structure formation, retrieving the one presented in the seminal paper of \citet{Heavens2003} where a product of the linear growth factors at different redshifts is present, acting on the matter power spectrum evaluated at the present time.

\begin{figure*}
	\begin{multicols}{2}
		\includegraphics[width=1.1\linewidth]{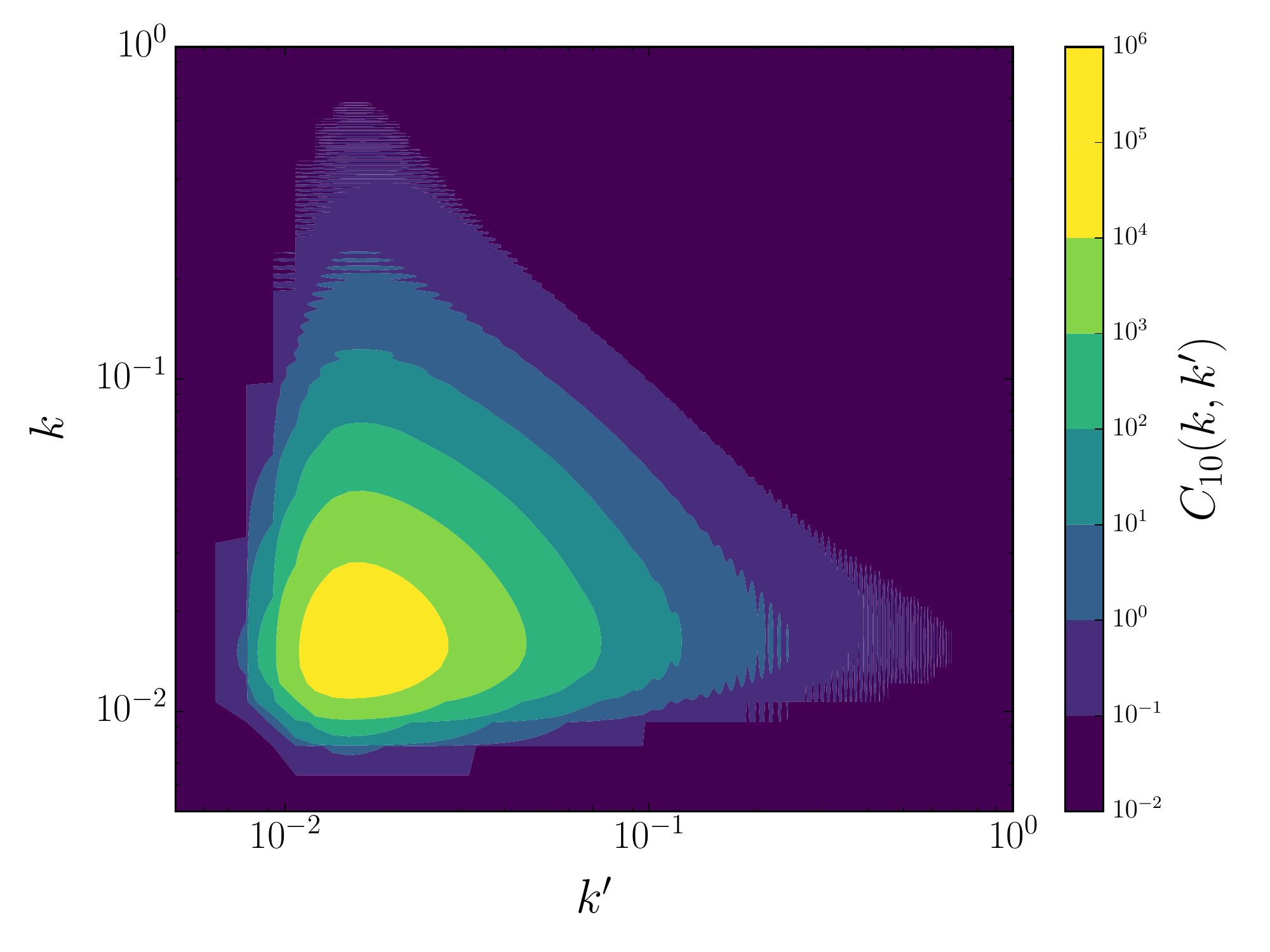}\par 
		\includegraphics[width=1.1\linewidth]{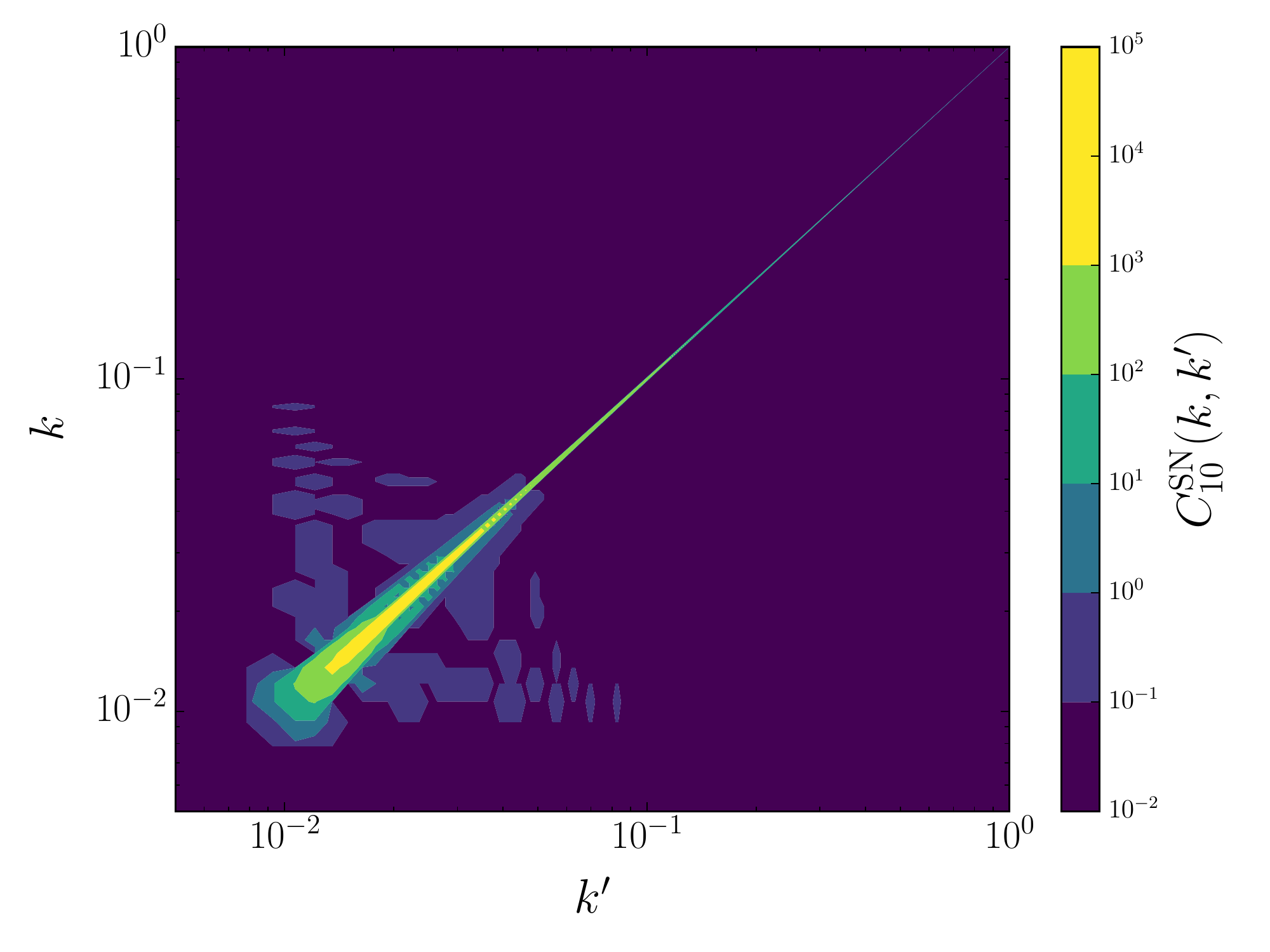}\par 
	\end{multicols}
	\begin{multicols}{2}
		\includegraphics[width=1.1\linewidth]{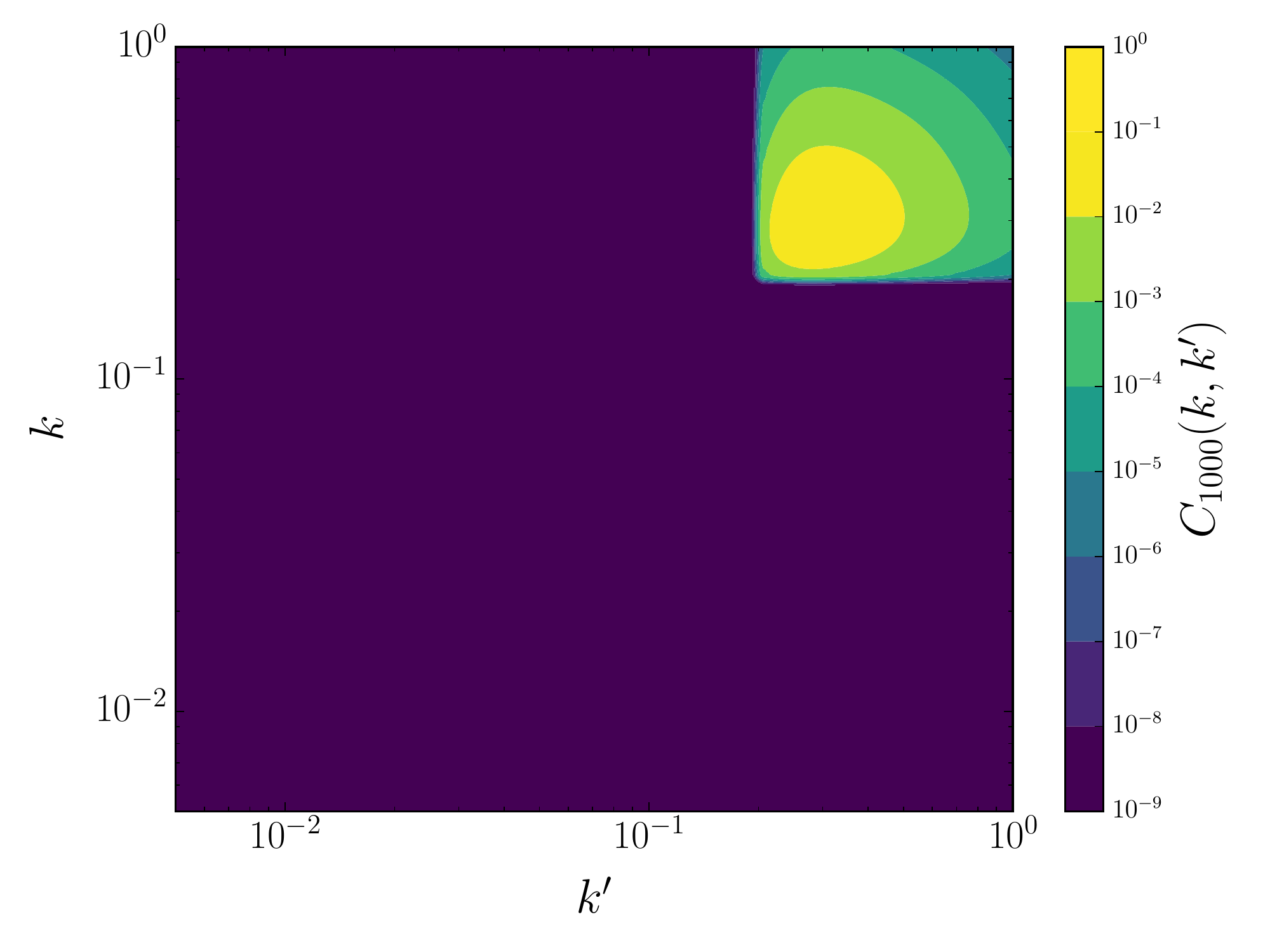}\par
		\includegraphics[width=1.1\linewidth]{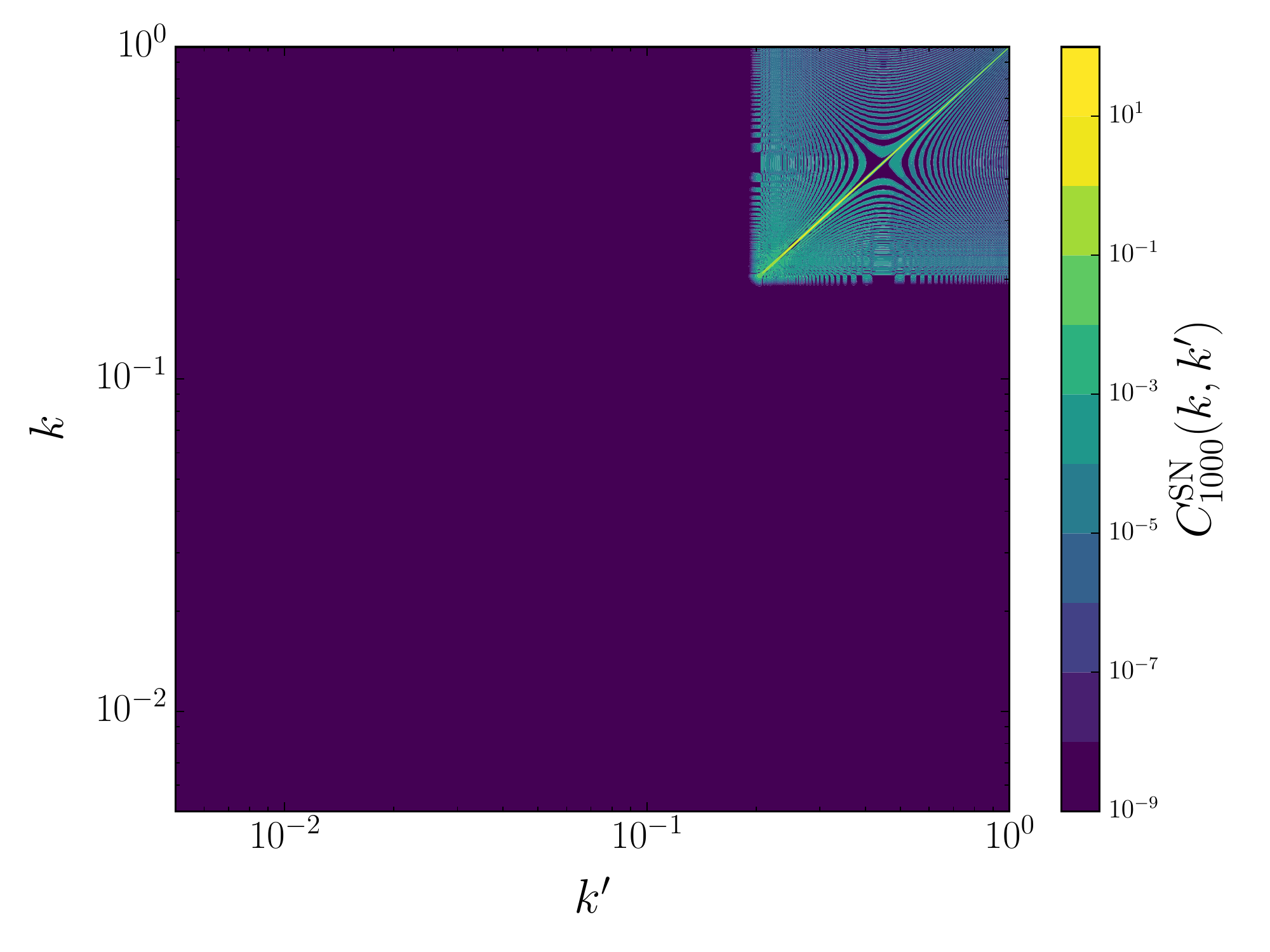}\par
	\end{multicols}
	\caption{Signal (\textit{left}, labelled $C_\ell$) and noise (\textit{right}, labelled $C_\ell^{\mathrm{SN}}$, where the subscript stands for shot noise) parts of the covariance matrix (Eqs.~\ref{eq:3Dcovariance} and~\ref{eq:noise}, respectively) for the minimum and maximum $\ell$-mode considered in the analysis, $\ell = 10$ (\textit{upper panels}) and $\ell = 1000$ (\textit{bottom panels}), respectively. Note the different ranges of the colour bars, in logarithmic scale. See also Fig.~\ref{fig:diagonal} for a comparison between the diagonal elements of the matrices, highlighting how different multipoles have contributions with different orders of magnitude and how the signal and noise part become dominant for low and high $\ell$ values, respectively.}
	\label{fig:covariances}
\end{figure*}

The noise term for the covariance matrix of the shear modes is given by the intrinsic ellipticity dispersion of source galaxies, as a result of the fact that the observed ellipticity $\epsilon$ is assumed to be the sum of the shear $\gamma$ and the intrinsic ellipticity $\epsilon_S$. The intrinsic ellipticity dispersion is given by $\left\langle \epsilon_S^2 \right\rangle = \sigma_\epsilon^2$. In the spherical Fourier-Bessel formalism, this gives
\begin{align}\label{eq:noise}
\left\langle \gamma_{\ell m}\left( k \right) \gamma_{\ell' m'}\left( k' \right) \right\rangle_{\mathrm{SN}} = \frac{\sigma_\epsilon^2}{2 \pi^2} \int \mathrm{d}z \, n_z(z) j_\ell \left[ k \chi_0(z) \right] j_{\ell'} \left[ k' \chi_0(z) \right] \delta_{\ell \ell'}^K \delta_{m m'}^K.
\end{align}
and we set $\sigma_\epsilon = 0.3$. This expression for the noise holds only in absence of intrinsic alignments, i.e. assuming that the intrinsic ellipticities of galaxies are uncorrelated \cite[see][for a study of intrinsic alignments in 3D weak lensing]{Merkel2013}.

\subsection{Tomography}\label{sec:3D2Dcompare}
Instead of keeping track of the photometric redshift error, as done in the 3D approach, by means of the probability $p(z_p|z)$ of estimating the redshift $z_p$ conditional on the true redshift $z$, another possibility is to assign every galaxy to a redshift bin. In this case, as opposed to Eq.~\ref{eq:3Dcovariance}, the flat sky tomographic cosmic shear power spectrum in tomographic bins $i$ and $j$ is given by
\begin{equation}\label{eq:2dtomo}
C^\kappa_{ij}(\ell)= \int\frac{\dd\chi}{\chi^2} W_i(\ell/\chi,\chi)W_j(\ell/\chi,\chi)\:P_\delta(\ell/\chi,\chi),
\end{equation}
where we used the Limber projection. The lensing efficiency function $W_i(\ell/\chi,\chi)$ is defined as
\begin{equation}\label{eq:weightfunction}
\begin{split}
W_i(\ell/\chi,\chi) = \frac{3\Omega_m}{4\chi_H^2}\int_\chi^\infty & \dd\chi^\prime\frac{\dd z}{\dd\chi^\prime}\frac{n_i(z(\chi^\prime))}{a(\chi')}\frac{\chi-\chi'}{\chi\chi'} \\ 
& \times\left(1+\frac{1}{\eta(\ell/\chi,\chi^\prime)}\right) \mu (\ell/\chi,\chi^\prime),
\end{split}
\end{equation}
$n_i(z(\chi))$ being the distance distribution of sources in the $i$-th bin which is normalised to one, $\int \dd\chi\:n_i\left(z(\chi)\right) = 1$.  Observed spectra suffer from Poissonian noise due to the intrinsic ellipticity dispersion of galaxies $\sigma_\epsilon$ and their finite number $n_0$. Choosing our tomographic bins so as to have equal number of galaxies in each of them, the observed tomographic weak lensing spectrum is given by
\begin{equation}\label{eq:tomo_s_plus_n}
\hat C^\kappa_{ij}(\ell) = C^\kappa_{ij}(\ell) + \frac{\sigma^2_\epsilon \, n_\mathrm{bins}}{n_0}\delta_{ij}.
\end{equation}
It should be noticed that comparing tomographic and 3D lensing must be done with some care since some approximations enter in Eq. (\ref{eq:2dtomo}). In particular, we made use of the flat sky approximation and the Limber projection \citep{Kaiser1992, Kaiser1998, Loverde2008}, neither of which is included in the 3D formalism. For a detailed discussion on this we refer to \citep{Kitching2016a, Kilbinger2017, Lemos2017} for an excellent discussion of various approximations performed in cosmic shear analyses.

\section{Methodology}
\label{sec:Fisher}

\begin{figure}
	\begin{center}
		\includegraphics[width = 0.45\textwidth]{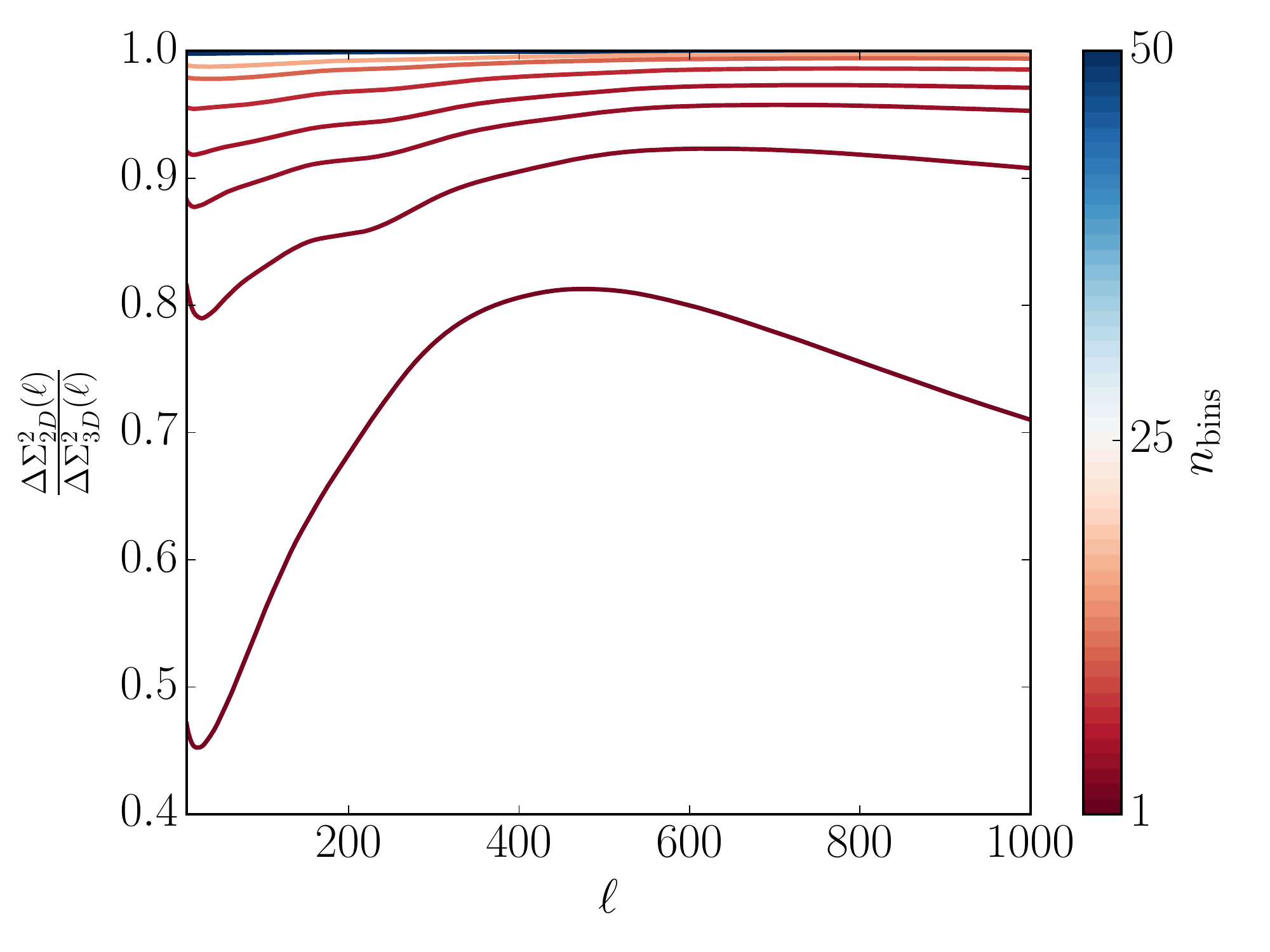}
		\caption{Differential signal-to-noise ratio of a tomographic analysis relative to a 3D analysis. The number of tomographic bins is shown in the colour bar.}
		\label{fig:SNR_3d_2d}
	\end{center}
\end{figure}

\begin{figure}
	\begin{center}
		\includegraphics[width = 0.45\textwidth]{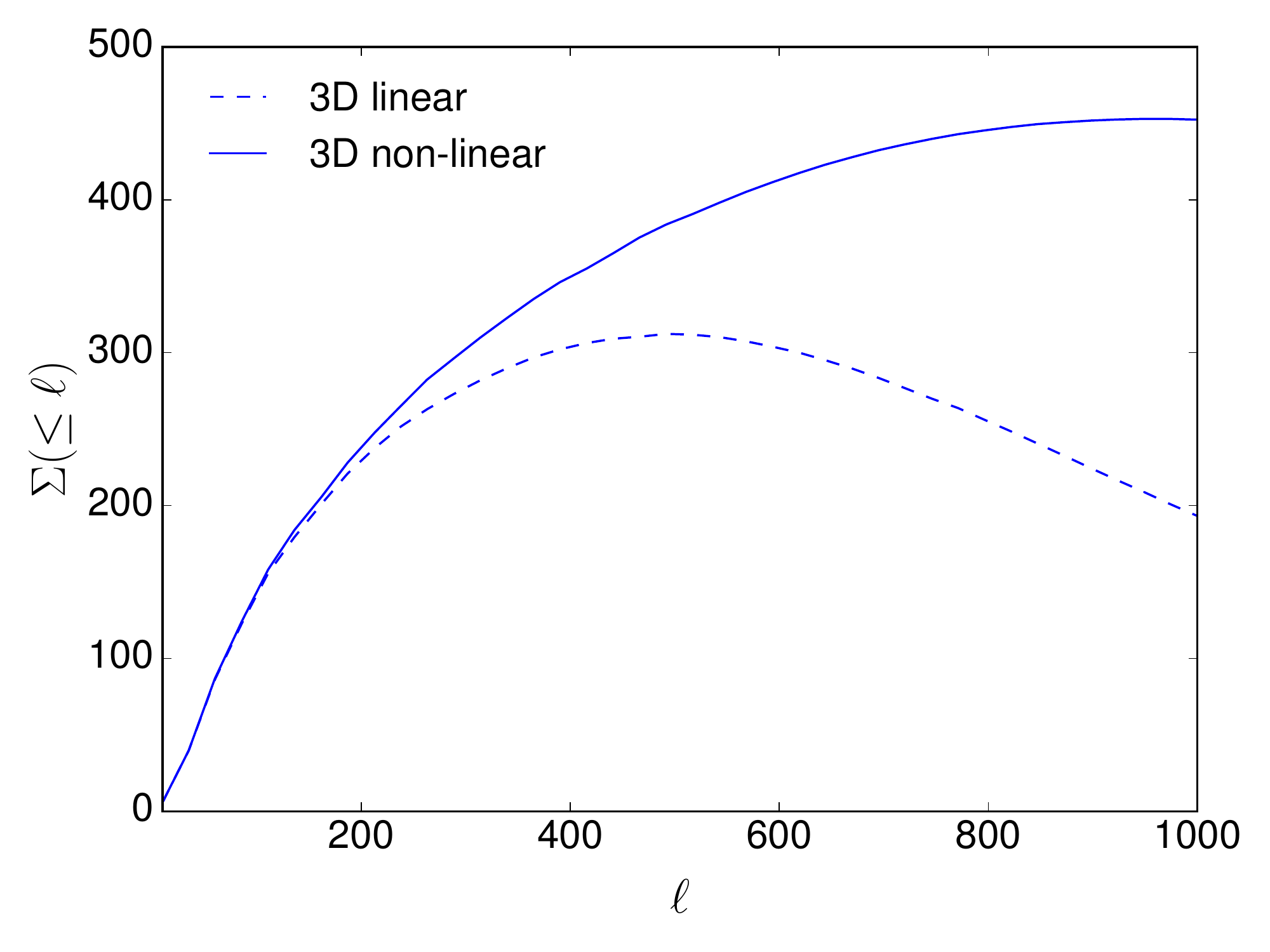}
		\caption{Differential signal-to-noise ratio, (\ref{eq:SNR}), i.e. $\Sigma^2$ gained at each multipole $\ell$: impact of the non-linear power on the 3D lensing signal. The solid curve shows the differential signal-to-noise with a non-linear power spectrum, while the dashed curve refers to a linear one. }
		\label{fig:SNR_3d_nonlin_lin}
	\end{center}
\end{figure}

\subsection{Fisher matrix forecasts}
In order to present forecasts for a Euclid-like experiment on the parameters considered, we perform a Fisher matrix analysis \citep{Tegmark1997}. Provided that the likelihood surface near the maximum is well approximated by a multivariate Gaussian, the Fisher matrix gives a realistic expectation of the foreseen error for a given experimental setting. The Fisher matrix is defined as the expectation value of the derivative of the logarithmic likelihood $\mathcal{L}$ with respect to the parameters $\theta_{\alpha}$:
\begin{align}\label{eq:Fisher_definition}
F_{\alpha \beta} \equiv  - \left\langle \frac{\partial^2 \mathrm{ln}\mathcal{L}}{\partial \theta^{\alpha} \partial \theta^{\beta}} \right\rangle,
\end{align}  
evaluated at the maximum of the log-likelihood $\mathcal{L}$, which in a forecast analysis coincides with the reference fiducial model. Once we have the Fisher matrix, the Cramer-Rao bound $\Delta \theta_{\alpha}^2 \geq (\boldsymbol{F}^{-1})_{\alpha \alpha}$ gives a lower limit on the expected marginal error on the parameter $\theta_{\alpha}$. If the data are Gaussian distributed and the mean values vanish, the Fisher matrix can be calculated from the covariance matrix and its derivatives with respect to the parameters \citep{Tegmark1997}
\begin{align}\label{eq:Tegmark}
F_{\alpha \beta} = \frac{1}{2} \mathrm{Tr} \left[ \boldsymbol{C}^{-1} \boldsymbol{C}_{,\alpha} \boldsymbol{C}^{-1} \boldsymbol{C}_{,\beta} \right].
\end{align}
where derivatives have been denoted with a comma.
Assuming full-sky coverage this expression can be simplified for a 3D weak lensing survey, as modes with different $\ell$ and $m$ are uncorrelated ($\delta^K_{\ell \ell'} \delta^K_{m m'}$ in eq.\eqref{eq:3Dcovariance}), leading to
\begin{align}\label{eq:Fishermatrix}
F_{\alpha \beta} = \frac{1}{2} \sum_\ell (2 \ell + 1) \mathrm{Tr}\left[ \boldsymbol{C}_\ell^{-1} \boldsymbol{C}_{\ell, \alpha} \boldsymbol{C}_\ell \boldsymbol{C}_{\ell, \beta} \right],
\end{align}
as there are $2\ell +1$ statistically independent $m-$ modes for each $\ell$. Note that expressions (\ref{eq:Tegmark}) and (\ref{eq:Fishermatrix}) are only exact if the data, in this case the modes $\gamma_{\ell m } (k)$, follow a Gaussian distribution. This is not the case for high $\ell$ values, where structures due to non-linear clustering dominate the lensing signal~\citep[for a discussion on non-Gaussian statistics of the weak lensing field see e.g.][]{Taruya2002, Joachimi2011, Clerkin2017}.  
However, for the purpose of this paper this assumption will not be of any harm since the basic parameter dependencies are captured well enough within this approximation.

\subsection{Observational effects and specifications}\label{subsec:observational}

A 3D weak lensing analysis depends crucially on redshift estimation of the source galaxies, which for next generation surveys like $Euclid$ \citep{EuclidRedBook2011} will be achieved using photometry, being the number of sources prohibitively high for spectroscopy. The estimated shear modes in \eqref{eq:3Dcovariance} keep into account two observational effects, which are inherent in a redshift survey. The first one is described by the quantity $G$ in \eqref{eq:G} and represents the distribution in redshift of the galaxies, mainly due to the fact that they become fainter as redshift increases.  For the source distribution we follow \citep{Amendola2016} and choose in \eqref{eq:G}
\begin{align}
n_z(z) \propto n_0 \left( \frac{\sqrt{2}}{z_m} \right) ^3 z^2 \exp \left[ - \left( \frac{\sqrt{2}z}{z_m}\right)^{3/2}\right],
\end{align}  
where $z_m$ is the median redshift of the survey and $n_0$ is the observed redshift-integrated source density $n_0$.
The second observational effect, kept into account in the quantity $F$ in \eqref{eq:F}, is the error associated to redshift estimation. This is described by the probability of estimating the reshift $z_p$ given the measured redshift $z$. We take this probability distribution to be a Gaussian
\begin{align}\label{eq:redshift_probability}
p(z_p|z) = \frac{1}{\sqrt{2 \pi} \sigma(z)} \exp \left[ - \frac{(z_p-z)^2}{2 \sigma^2(z)} \right],
\end{align}
with a redshift-dependent dispersion 
\begin{align}
\sigma(z) = \sigma_z (1+z).
\end{align}
If the sky coverage is not complete Eq. (\ref{eq:Fishermatrix}) is not completely correct, since the spherical basis is no longer orthogonal. Nonetheless, Eq. (\ref{eq:Fishermatrix}) is a good approximation by just multiplying the right hand side with the sky-fraction $f_\mathrm{sky}=\Omega_\mathrm{survey}/\Omega_\mathrm{sky}$. The choice we make for the specifications used in the analysis is summarized in Table \ref{tab:specifications}.

\subsection{Scales considered and non-linear corrections}\label{subsec:scales}

\begin{figure}
\begin{multicols}{2}
    \includegraphics[width=1.1\linewidth]{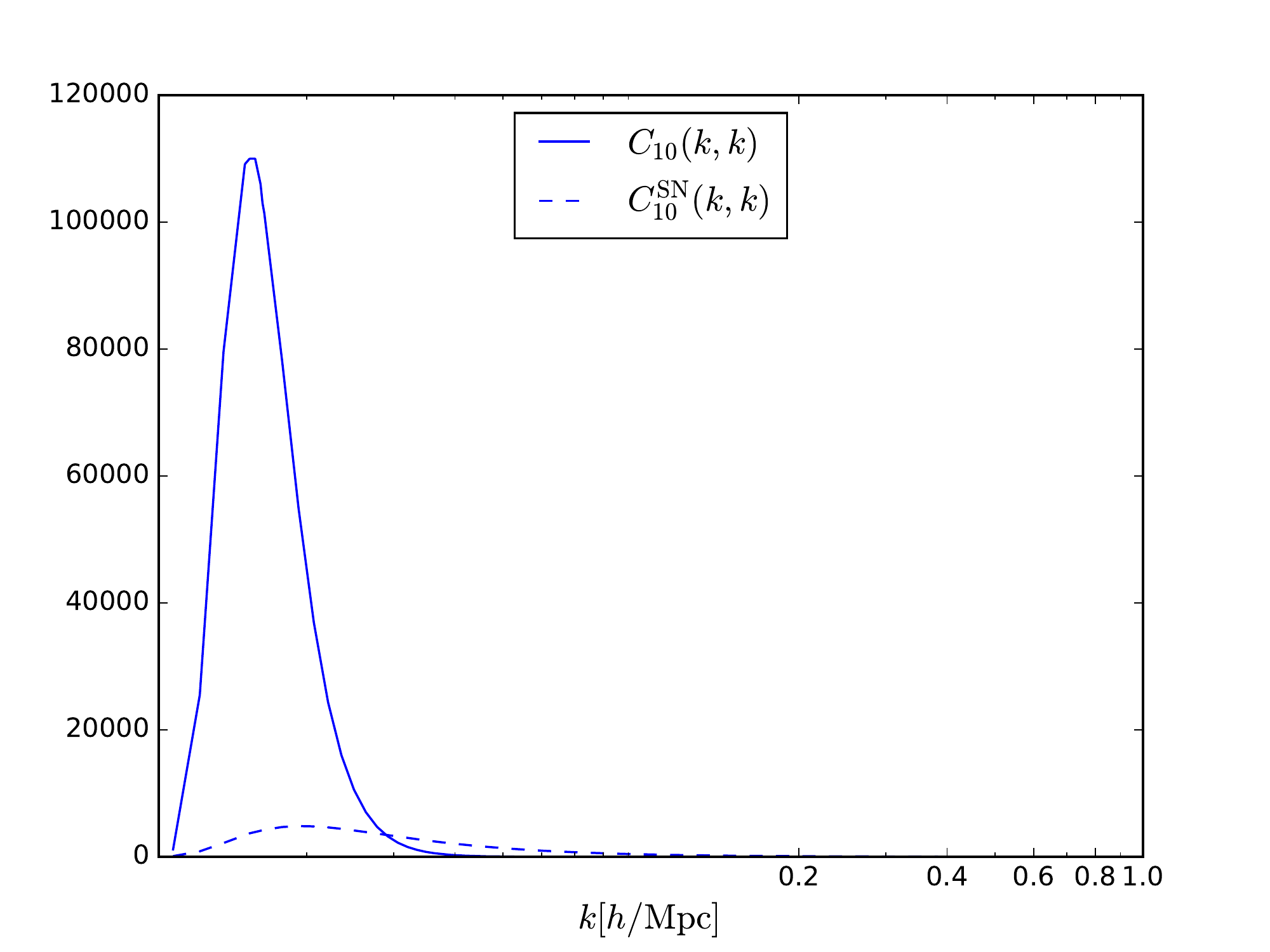}\par 
    \includegraphics[width=1.1\linewidth]{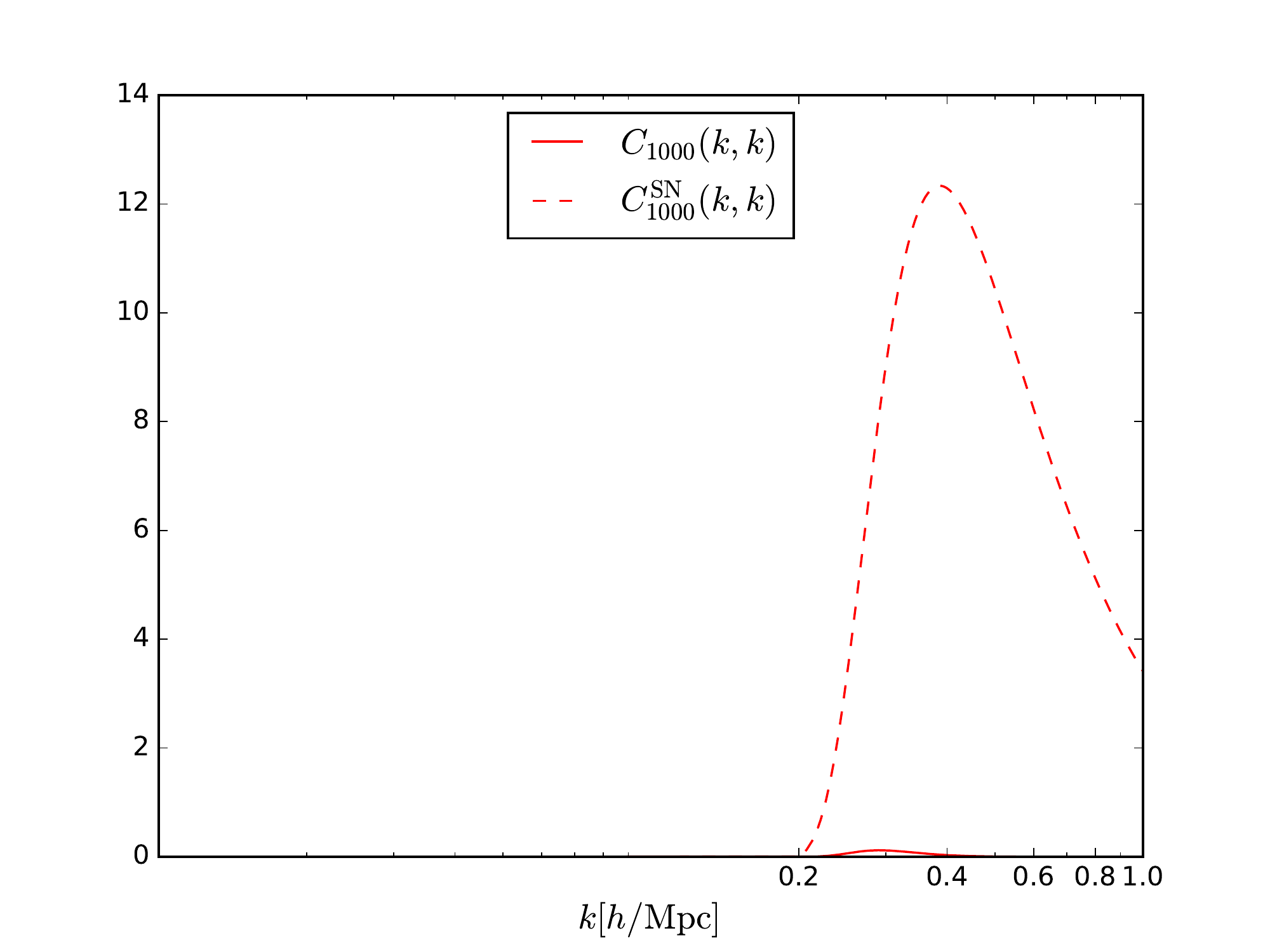}\par 
    \end{multicols}
\caption{Comparison between the diagonal elements of the signal (\textit{solid line}, as given by eq.\ref{eq:3Dcovariance} and labelled $C_\ell$) and noise (\textit{dashed line}, as given by eq.\ref{eq:noise} and labelled $C_\ell^{\mathrm{SN}}$, where the subscript stands for shot noise) contributions to the covariance matrices of the shear modes, for the minimum and maximum angular multipole considered in this analysis, i.e. $\ell = 10$ (\textit{blue}) and $\ell = 1000$ (\textit{red}), respectively. Note the different orders of magnitude for the different multipoles, and how the signal prevails on the noise for low multipoles, while the noise dominates for higher $\ell$ values. Note also the log-scale on the $x$-axis, to help identify the different $k$-regions where most of the contributions come from, for different multipoles.}
\label{fig:diagonal}
\end{figure}

The cuts in angular and radial scales that we perform, $\ell_{\mathrm{max}} = 1000$ and $k_{\mathrm{max}}= 1.0 \, h/ \mathrm{Mpc}$, are such that we avoid the deeply non-linear regime of structure growth. We demonstrate this point in \autoref{fig:covariances}, showing the signal and noise parts of the covariance matrices (Eq.~\ref{eq:3Dcovariance} and Eq.~\ref{eq:noise}) for two $\ell$-modes, 10 and 1000, which correspond to the mininum and maximum multipole considered in our analysis, respectively. We notice how even for the higher $\ell$ case, the range of $k$ scales considered justifies our choice to use the linear power spectrum for our analysis, since the higher-$k$ part of the spectrum is dominated by the noise (notice the different orders of magnitude between the signal and noise contibutions). In Fig.~\ref{fig:diagonal} we plot only the diagonal contributions to the covariance matrices, distinguishing between the signal and noise parts, for $\ell = 10$ and $\ell =1000$, our minimum and maximum angular multipoles. One can see how the orders of magnitude of the covariance matrices between different multipoles change and also how the dominance of the signal over the noise part gets inverted going from low to high $\ell$ values.

We compare forecasts obtained with a linear matter power spectrum in the calculation of the shear covariances to those obtained with a non-linear power spectrum. The current lack of solid understanding for non-linear corrections, in $\Lambda$CDM and even more in a modified gravity context, implies that any non-linear prescription should be employed with caution. The matter power spectra are produced using the \texttt{hi\char`_class} code \citep{Zumalacarregui2016}, a modification of the \textsc{Class} Boltzmann solver \citep{Lesgourgues2011} for Horndeski theories of gravity. In particular, \texttt{hi\char`_class} allows the user to choose the parameterization for the $\alpha(\tau)$ functions which traces the evolution of the dark energy component, and the code takes then as input the proportionality coefficients $\hat{\alpha}$ (Eq. \ref{eq:alpha_time_evolution}). The choice for the fiducial values of $\hat{\alpha}_B$ and $\hat{\alpha}_M$, reported in \autoref{tab:parameters_ak001_tomo3Dlin}, is close enough to $\Lambda$CDM to represent General Relativity with an additional cosmological constant, without incurring in numerical difficulties in \texttt{hi\char`_class} if the $\hat{\alpha}$ coefficients are all set to zero exactly. Since \texttt{hi\char`_class} is a linear code it produces linearly evolved power spectra only. Non-linear corrections can however be incorporated by applying a non-linear transfer function using \texttt{halofit} \citep{Smith2003,Takahashi2012,Bird2012} as implemented in \texttt{hi\char`_class} or a more state of the art version, \textsc{HMcode}, developed by \citet{Mead2015}, which we employ for our non-linear forecasts. Both \texttt{halofit} and \textsc{HMcode} however deal with non-linearities only in a setting where standard General Relativity is true. In order to get consistent constraints we therefore follow \citet{Alonso2016} and introduce a screening mechanism to recover General Relativity on small scales by a phenomenological modification of the $\alpha(\tau)$ functions, employing a Gaussian kernel in Fourier space with a characteristic scale $k_V$:
\begin{align}\label{eq:screening}
\alpha(\tau) \rightarrow \alpha(\tau, k) = \alpha(\tau) \exp \left( -\frac{1}{2} \left( \frac{k}{k_V}\right)^2 \right)\; .
\end{align}
We marginalise over the scale $k_V = 0.1 \, h/ \mathrm{Mpc}$ at which the screening mechanism becomes effective. The fiducial choice for the screening is important in the sense that non-linear effects become important at scales smaller than $0.1\, h/\mathrm{Mpc}$. Additionally,  \citet{Barreira2013} showed that the typical scale of Vainshtein screening is roughly at $0.1\, h/\mathrm{Mpc}$. A plot with two different choices of $k_V$ can be found in Appendix~\ref{app:kv}.

\section{Results}\label{sec:selected_results}
In this section we will investigate the signal strength of a weak lensing analysis carried out using the full photometric redshift information via the 3D method, as well as by using a tomographic technique. As already mentioned we calculate the tomographic lensing power spectrum using the Limber approximation; for a more detailed discussion we refer to \citet{Kitching2016a}. We then show the possible constraints on Horndeski cosmological models with survey specifications given in \autoref{tab:specifications}.

\subsection{Signal to noise for 3D and tomographic weak lensing}
\autoref{fig:SNR_3d_2d} shows the differential signal-to-noise (SNR) curve for a tomographic survey relative to a 3D analysis as a function of the number of tomographic bins. The total signal-to-noise ratio is calculated as
\begin{equation}\label{eq:SNR}
\Sigma^2(\leq\ell) = f_\mathrm{sky}\sum_{\ell' = \ell_\mathrm{min}}^{\ell}\frac{2\ell' +1 }{2} \mathrm{Tr}\left[\boldsymbol{C}^{-1}_{\ell'}\boldsymbol{S}_{\ell'}\boldsymbol{C}_{\ell'}^{-1}\boldsymbol{S}_{\ell'}\right]\equiv \sum_{\ell' = \ell_\mathrm{min}}^{\ell}\Delta\Sigma^2(\ell')\; ,
\end{equation}
where $\boldsymbol{S}$ is the signal covariance (\ref{eq:3Dcovariance}) or (\ref{eq:2dtomo}) only, while $\boldsymbol{C}$ refers to the sum of signal and noise, i.e. (\ref{eq:3Dcovariance})$+$(\ref{eq:noise}) or (\ref{eq:tomo_s_plus_n}).

\begin{table*}
\begin{center}
 \begin{tabular}{llcccccccccccc}
  \hline
  &  & $\hat\alpha_{K}$  & $\hat\alpha_\mathrm{B}$ &   $\hat\alpha_\mathrm{M}$ & $\hat\alpha_{T}$ & $\Omega_\mathrm{m}$ & $\sigma_8$ & $h$ & $n_\mathrm{s}$ & $\Omega_\mathrm{b}$ & $\sum m_\nu$\\
  & fiducial value  & 0.01 & 0.05 & 0.05 & 0.05 & 0.314 & 0.834 & 0.678 & 0.968 & 0.0486 & 0.05 \\
  \hline \hline
 & \specialcell{error 3DWL  linear, \\ varying $\hat\alpha_{K}$ (fiducial value 0.01), \\ varying $\hat\alpha_{T}$ (fiducial value 0.05)} & \specialcell{137\\(13564$\%$)} & \specialcell{0.82\\(78$\%$)} & \specialcell{2.24\\(213$\%$)} & \specialcell{5.62\\(535$\%$)} & \specialcell{0.015\\(4.78$\%$)} & \specialcell{0.045\\(5.39$\%$)} & \specialcell{0.307\\(45.28$\%$)} & \specialcell{0.101\\(10.43$\%$)} & \specialcell{0.027\\(55.55$\%$)} & \specialcell{0.484\\(968$\%$)} \\ 
  \hline 
 & \specialcell{error tomography  linear, \\ $\hat\alpha_{K} = 0.01$, $\hat\alpha_{T} = 0$} & / & \specialcell{0.57\\ (54$\%$)} & \specialcell{1.66\\(158$\%$)} & / & \specialcell{0.017\\(5.41$\%$)} & \specialcell{0.056\\(6.71$\%$)} & \specialcell{0.257\\(37.90$\%$)} & \specialcell{0.088\\(9.09$\%$)} & \specialcell{0.022\\(45.27$\%$)} & \specialcell{0.472\\(944$\%$)} \\
 \hline   
  & \specialcell{error 3DWL linear,\\ $\hat\alpha_{K} = 0.01$, $\hat\alpha_{T} = 0$} & / & \specialcell{0.43\\(41$\%$)} & \specialcell{1.32\\(126$\%$)} & / & \specialcell{0.013\\(4.14$\%$)} & \specialcell{0.042\\(5.03$\%$)} & \specialcell{0.239\\(35.25$\%$)} & \specialcell{0.080\\(8.26$\%$)} & \specialcell{0.021\\(43.21$\%$)} & \specialcell{0.408\\(816$\%$)} \\
  \hline
  & \specialcell{error 3DWL linear,\\ $\hat{\alpha}_K = 10$, $\hat\alpha_{T} = 0$} & / & \specialcell{0.48\\(46$\%$)} & \specialcell{1.40\\(133$\%$)} & / & \specialcell{0.014\\(4.46$\%$)} & \specialcell{0.043\\(5.15$\%$)} & \specialcell{0.238\\(35.10$\%$)} & \specialcell{0.080\\(8.26$\%$)} & \specialcell{0.021\\(43.21$\%$)} & \specialcell{0.400\\(800$\%$)}\\
  \hline
 & \specialcell{error 3DWL non-linear,\\ $\hat{\alpha}_K = 0.01$, $\hat\alpha_{T} = 0$}  & / & \specialcell{0.25\\(24$\%$)} & \specialcell{0.55\\(52$\%$)} & / & \specialcell{0.011\\(3.50$\%$)} & \specialcell{0.010\\(1.20$\%$)} & \specialcell{0.134\\(19.76$\%$)} & \specialcell{0.038\\(3.92$\%$)} & \specialcell{0.016\\(32.92$\%$)} & \specialcell{0.229\\(458$\%$)} \\
 \hline
 \end{tabular}
\captionof{table}{Marginalized errors $\sigma_i = \left[(\boldsymbol{F}^{-1})_{ii}\right]^{1/2}$, for the survey characteristics from \autoref{tab:specifications}. We compare a tomographic analysis with the 3D analysis using a linear power spectrum. Furthermore the influence of the value $\hat\alpha_K$ is investigated. Lastly the impact of non-linear clustering is quantified. When reporting the relative percentage error for the $\hat\alpha$ coefficients, we calculate it with respect to their fiducial values increased by one, as they are indeed expected to be $O(1)$ if one assumes the parameterization tracing the dark energy density fraction to model the time evolution of the $\hat\alpha(\tau)$ functions \citep{Bellini2016}.}
 \label{tab:parameters_ak001_tomo3Dlin}
\end{center}
\end{table*}

The number of tomographic redshift bins is shown in the colour bar. Clearly an increase of the number of bins used increases the SNR, however, the gain in signal saturates for $n_\mathrm{bins} \approx 15$ due to the non-vanishing cross-correlation between the different bins. It should be noted that there is in principle an additional effect due to the finite width of the photometric redshift estimation. If the average bin width in the tomographic case is of the order of the width of the distribution of redshift estimation error, the correlation between neighbouring bins will be underestimated, thus producing artificially signal. This effect is, however, very small as long as $\sigma_z$ is sufficiently small.
In the 3D case this correlation is represented by the covariance (\ref{eq:3Dcovariance}).

\begin{figure}
	\begin{center}
		\includegraphics[width = 0.46\textwidth]{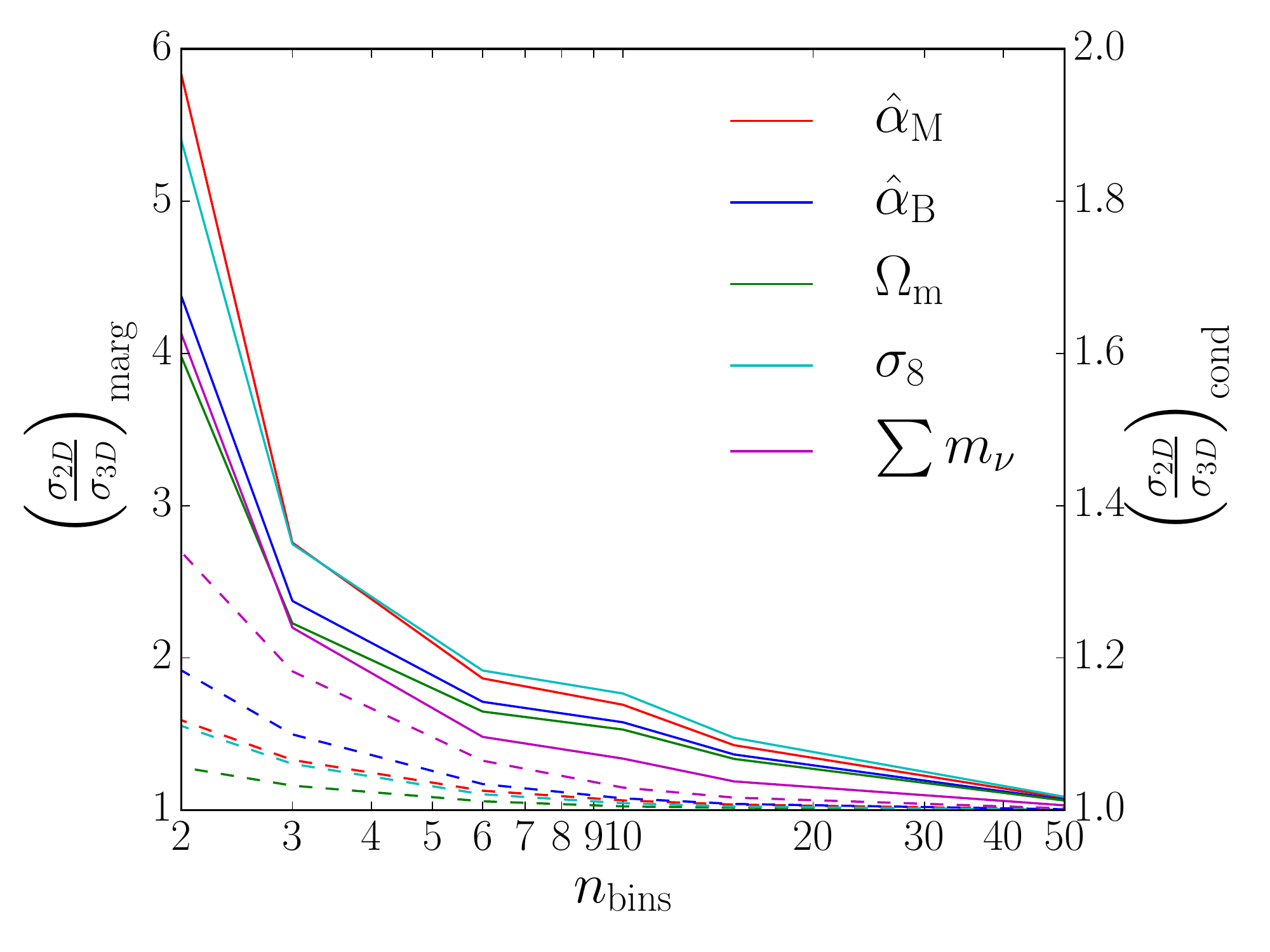}
		\caption{Errors on different parameters from a tomographic analysis relative to a 3D analysis as a function of the number of tomographic bins, $n_\mathrm{bins}$ (in log-scale on the $x$-axis). The solid lines show the ratio of the marginal errors belonging to the left $y$-axis, while the dashed lines show the conditional errors belonging to the right $y$-axis.}
		\label{fig:error_tomo_3d}
	\end{center}
\end{figure}

\autoref{fig:SNR_3d_nonlin_lin} displays the impact of non-linear clustering: if one only considers linear structure growth (dashed line), the shot noise starts dominating the signal at $\ell\approx 450$. For the non-linear power spectrum instead the differential signal-to-noise rises until $\ell\approx 1000$ (solid line) due to the enhancement of small scale structure by non-linear clustering. This shows the importance of the inclusion of high multipoles into the analysis. More specifically we see that the non-linear effects become important already at a relatively low $\ell \lesssim 200$.

\subsection{Cosmological constraints on Horndeski functions}
In our forecasts we fix $\alpha_T$ very close to zero and do not consider it as a parameter in our Fisher matrix analysis, reflecting the recent very strong constraints on the gravitational waves speed set by the detection of the binary neutron star merger GW170817 and the gamma ray burst GRB170817A \citep{Abbott2017b, Abbott2017a, Baker2017, Creminelli2017, Ezquiaga2017, Sakstein2017}. 
Furthermore the kineticity $\alpha_K$ is largely unconstrained by cosmological observables \citep{Bellini2016, Alonso2016}, therefore we fix the coefficient $\hat{\alpha}_K$ to its fiducial value. However, we study the impact of the choice of $\hat\alpha_K$ on the constraints on the other parameters by choosing two values which differ by three orders of magnitude.
\begin{figure*}
\centering
\includegraphics[width = \textwidth]{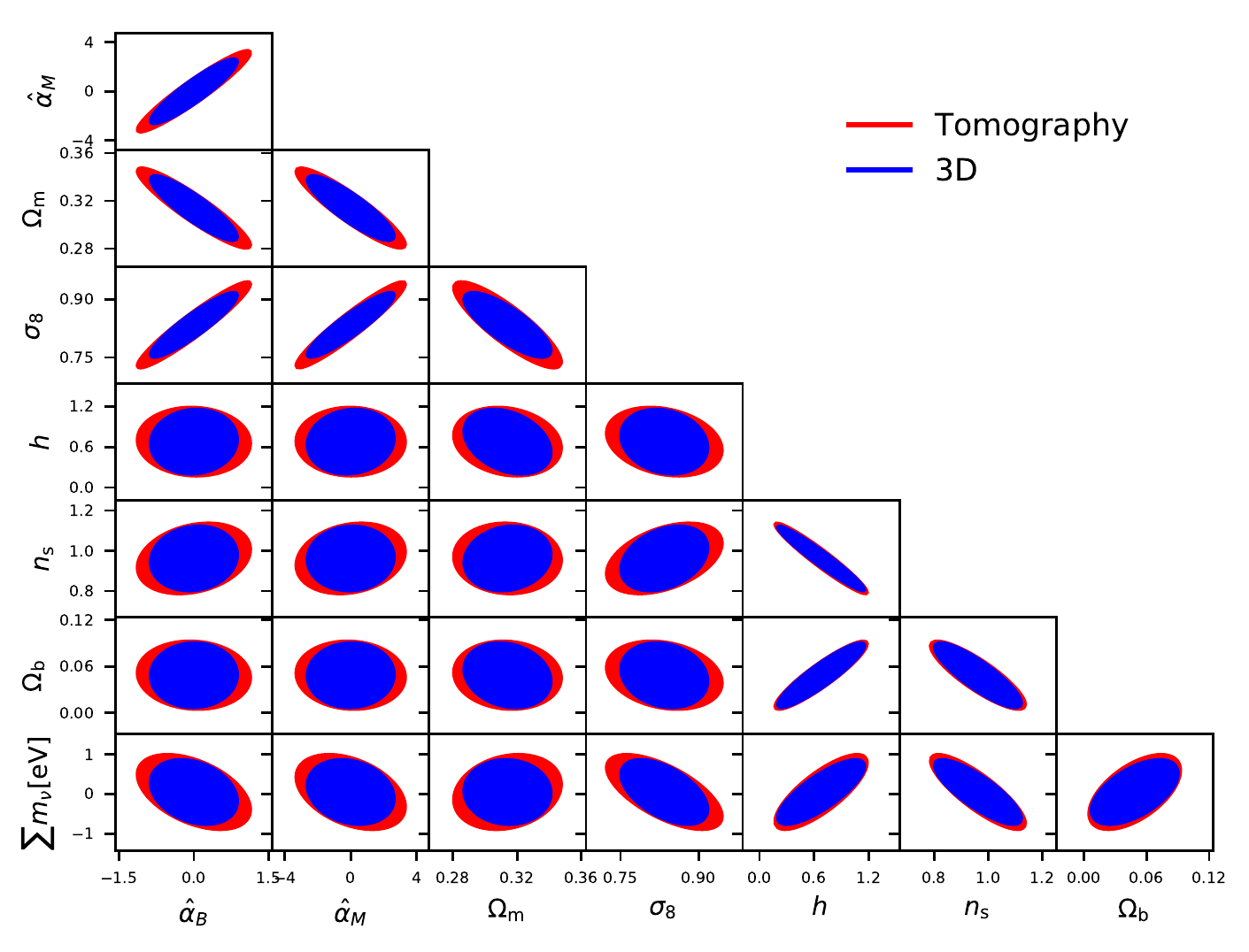}
\vspace*{-0.8cm}
\caption{1-$\sigma$ forecast contours for a Euclid-like survey, showing a comparison between a fully 3D (\textit{blue}) and a tomographic cosmic shear analysis (\textit{red}). The fiducial values can be found in \autoref{tab:parameters_ak001_tomo3Dlin}, while $\hat{\alpha}_K = 0.01$ and the survey specifications are given in \autoref{tab:specifications}. We used the linear matter power spectrum for both.}\label{fig:Comparison_3dtomo_lin_ak001}
\end{figure*}
As already seen in \autoref{fig:SNR_3d_2d} a lot of signal comes from non-linear scales, in fact for Euclid one expects that about two thirds of the total signal to noise, $\Sigma(<\ell_\mathrm{max})$ with $\ell_\mathrm{max} \approx 2000$, originate from non-linear scales. It is therefore evident that one has to include non-linear clustering in the analysis in order to get the necessary statistical power to constrain a high-dimensional parameter space.

As seen before, increasing the number of tomographic bins yields more signal. In the inference process, however, the sensitivity to the model parameters plays an important role. For linear model parameters one expects the sensitivity to be a rescaled version of the SNR curve. In \autoref{fig:error_tomo_3d} we show the marginal and conditional errors of a tomographic analysis relative to the 3D analysis for a few parameters. The marginal errors are more strongly affected, since the contributions from the conditional errors add up during the marginalization procedure. Furthermore, we see the same trend as for the SNR: the expected errors tend towards the errors of a 3D analysis for $n_\mathrm{bins}\gg 1$.

\autoref{fig:Comparison_3dtomo_lin_ak001} shows a comparison between cosmological constraints obtained with 3D cosmic shear and tomography with specifications from \autoref{tab:specifications}. Constraints from a 3D analysis are tighter than those from tomography, due to the increased redshift information. Furthermore, the degeneracies are in all cases very similar for the two methods, which is expected since the two methods probe the same quantity. In particular we find the usual degeneracy in $\Omega_\mathrm{m}$ and $\sigma_8$, which is slightly reduced in the 3D case. Generally the biggest improvement can be seen for parameters carrying information about the background evolution and the growth of structures; in contrast, parameters such as the spectral index $n_\mathrm{s}$ are not that much influenced.

\begin{figure*}
\begin{center}
\includegraphics[width = \textwidth]{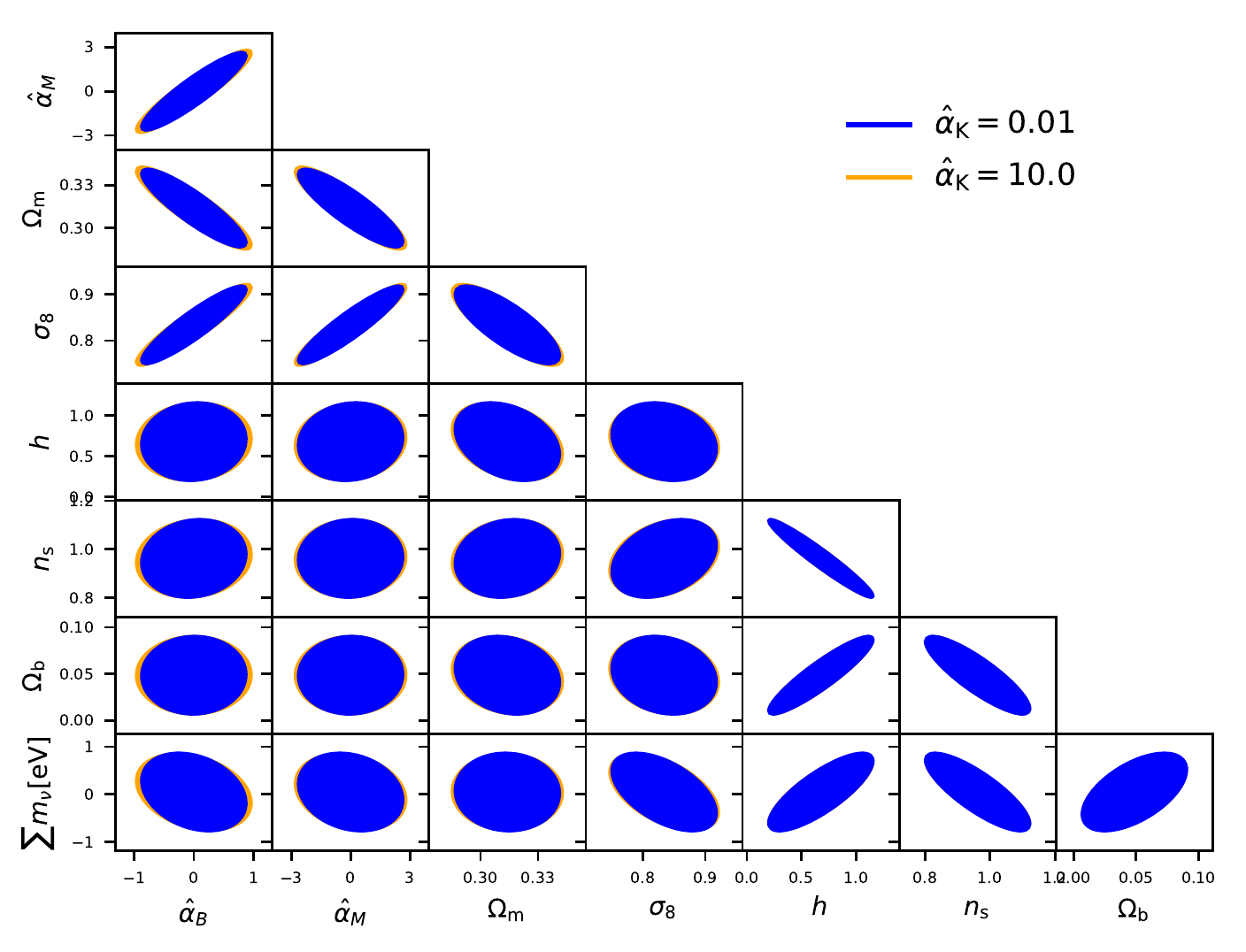}
\vspace*{-0.8cm}
\caption{Impact of the choice of $\hat\alpha_K$ for a 3D cosmic shear analysis with fiducial values from \autoref{tab:parameters_ak001_tomo3Dlin} and survey specifications given in \autoref{tab:specifications}. We used a linear power spectrum for the analysis and show the difference in the 1-$\sigma$ contours when fixing $\hat\alpha_K = 0.01$ (\textit{blue}) or $\hat\alpha_K = 10$ (\textit{orange}).}
\label{fig:3D_alphak_compare}
\end{center}
\end{figure*}

In \autoref{fig:3D_alphak_compare}, instead, we study the impact of the choice of $\hat\alpha_K$ and compare constraints obtained only with 3D cosmic shear using a linear matter power spectrum, but with two different choices of fixed $\hat{\alpha}_K$, namely $\hat{\alpha}_K = 0.01$ and $\hat{\alpha}_K = 10$. We find, in agreement with \citet{Alonso2016}, that the choice of $\hat{\alpha}_K$ does not affect the large-scale structure observables significantly. Since the largest effect on structure formation of $\hat\alpha_K$ comes from very large scales beyond those considered in this work, we do not expect any significant dependence of the Fisher matrix on this parameter.

\begin{figure*}
\begin{center}
\includegraphics[width = \textwidth]{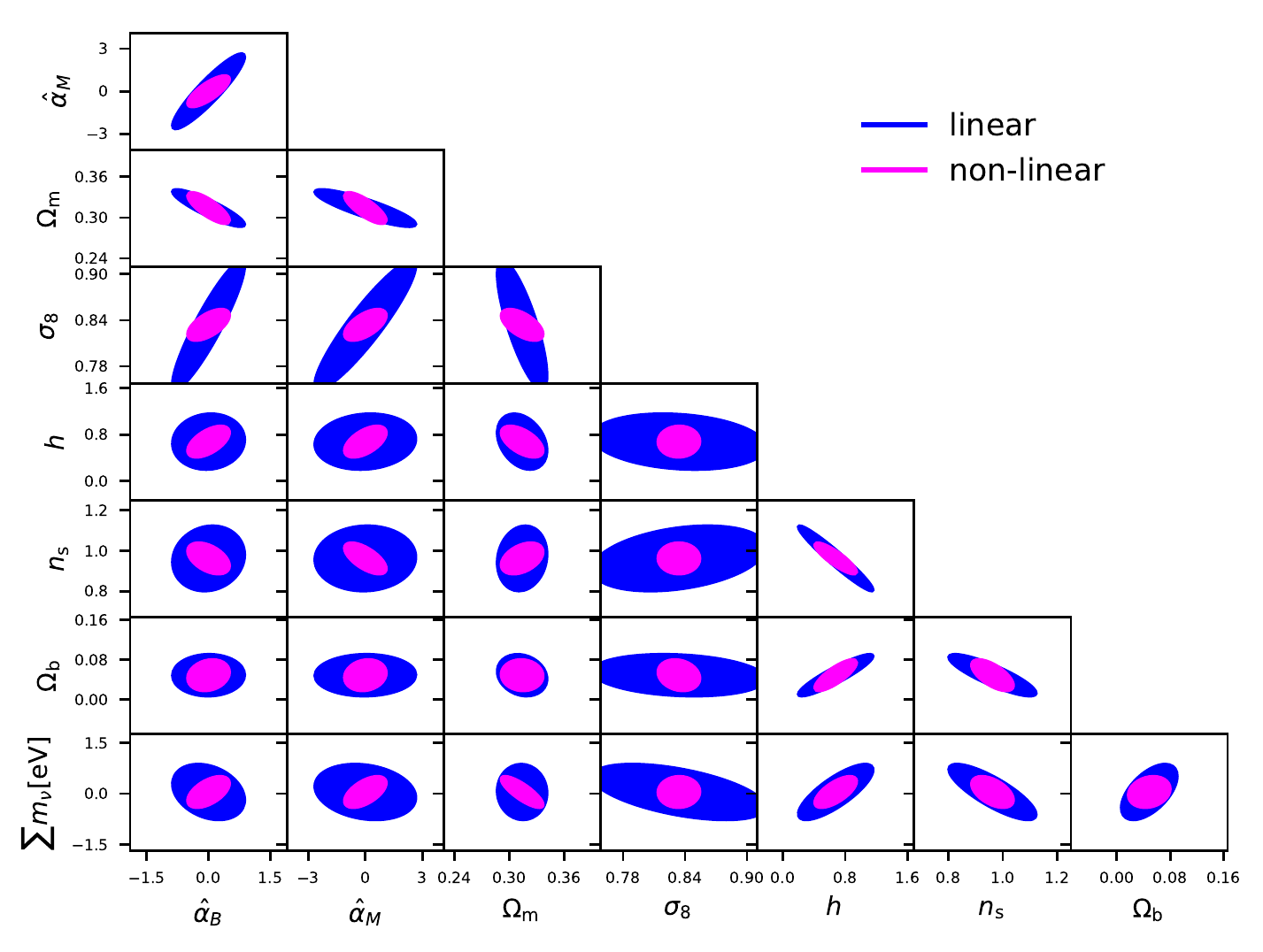}
\caption{Impact of non-linear clustering with fiducial values from \autoref{tab:parameters_ak001_tomo3Dlin} and survey specifications are given in \autoref{tab:specifications}. We show the constraints obtained with the linear power spectrum in \textit{blue} and with the non-linear one in \textit{magenta}.}
\label{fig:Comparison_3d_linnonlin_ak001}
\end{center}
\end{figure*}

Finally we investigate the impact of non-linear clustering in \autoref{fig:Comparison_3d_linnonlin_ak001} as outlined before.  Constraints are, as expected, tighter with the addition of the non-linear corrections.
In particular we find a significant gain in $\Omega_\mathrm{m}$, $\sigma_8$ and $\sum m_\nu [\mathrm{eV}]$. Other parameters such as the spectral index $n_\mathrm{s}$ and the Hubble constant $h$ are not that much affected, since the main characteristics are already captured in the linear power spectrum. Furthermore we find a gain in sensitivity in $\hat\alpha_B$ and $\hat\alpha_M$. This reduction of the error for the modified gravity parameters is mainly due to the marginalisation process and the degeneracies with the other parameters such as $\Omega_\mathrm{m}$, which are better constrained now.
In fact, it should be noted that the conditional constraints on $\hat\alpha_\mathrm{M}$ and $\hat\alpha_\mathrm{B}$ become slightly worse then in the linear case, which has a subtle reason: the screening scale is chosen such that modified gravity effects are suppressed as soon as non-linear effects set in, on the other hand however, there is loss of power on intermediate scales. This effectively yields a loss in sensitivity, since the full effect of modified gravity is only present up to intermediate scales, whereas the signal gain on small scales does not contribute to the Fisher matrix. Finally, as a consequence of the screening mechanism, the orientation of the ellipses, for example in the case $\Omega_m - \hat{\alpha}_M$ and $\sigma_8 - \hat{\alpha}_M$, can change; this is due to the increase in signal at high $\ell$ values and the change of the sensitivity to cosmological parameters, especially because the sensitivity to the modified gravity parameters on those scales vanishes by construction.  

In order to exploit the full potential of non-linear clustering one would need to have a reliable model for non-linear structure formation in a modified gravity setting. The way it is presented here effectively assumes that modifications of the gravitational field equation only play a role at linear order, while higher orders are treated in the usual framework of perturbation theory in a $\Lambda$CDM cosmology.

\section{Discussion and conclusions}\label{sec:conclusions}

In this work we investigated the performance of a 3D analysis of cosmic shear measurements as a probe of Horndeski theories of modified gravity. We set constraints by means of a Fisher matrix analysis on a set of parameters that completely describe the evolution of linear perturbations in Horndeski gravity, using the specifications of a future Euclid-like experiment. We placed simultaneously our constraints on both the modified gravity parameters and on a set of standard cosmological parameters, including the sum of neutrino masses. Analogous forecasts for a tomographic analysis with six bins were produced given the same specifications of the cosmic shear experiment, with the aim of comparing the two methods. Our analysis was restricted to angular modes $\ell \leq 1000$ and $k \leq 1 \, h/\mathrm{Mpc}$, to avoid the deeply non-linear regime of structure growth. We summarize our results as follows.
  
The signal-to-noise ratio of both a 3D analysis and a tomographic one is very similar, since it is mainly driven by the amplitude of the lensing signal and a tomographic method effectively agrees with a decomposition into spherical harmonics and radial Bessel functions if the bin width gets as small as the width of the photometric redshift errors. 

3D cosmic shear provides tighter constraints than 10 bins tomography. Even with our conservative cut in angular and radial scales and using a linear matter power spectrum for the calculation of the covariance of the shear modes, 3D weak lensing performs better than tomography for all cosmological parameters, with both methods showing very similar degeneracies. For the parameters of the \citet{BelliniSawicki2014} parametrization describing Horndeski theories, the gain is of the order of roughly 20 $\%$ in the errors.

We investigated the impact of the fiducial value chosen for the kineticity and found that the constraints are largely unaffected by the choice of $\hat\alpha_K$. In particular we used $\hat\alpha_K  =0.01$ and $\alpha_K = 10$.

To illustrate the importance of non-linear corrections, we showed the expected improvement in the size of the constraints obtained employing a non-linear matter power spectrum: the results obtained in this case serve as an illustrative example of the constraining power of non-linear scales. In order to obtain a complete and self-consistent picture, one would need a formalism to construct the non-linear corrections in a general modified gravity setting \citep[see e.g.][]{Lombriser2016, Fasiello2017}. Here we introduced an artificial screening scale, which pushes the deviations from General Relativity to zero below its value. This is however not a fully exhaustive ansatz and many more investigations in this direction are required.
The gain in signal if non-linear clustering is considered clearly shows the importance and calls for the development of analytic or semi-analytic prescriptions for the treatment of non-linear scales in $\Lambda$CDM and modified gravity. These will play a crucial role in allowing cosmic shear measurements to set strong constraints on parameters describing deviations from General Relativity. Due to the screening the constraints on modified gravity parameters are only improved because of the marginalization over the remaining parameters. 

Compared to the analysis of \citet{Pratten2016}, our study extends the scope to the full Horndeski class and we do not fix all the parameters describing the background to their $\Lambda$CDM values. This large parameter space and the fact that we only considered weak gravitational lensing as our observable makes our constraints less tight than the ones presented in \citet{Alonso2016}, given also that our range in scales is less extended. Additionally, we present a 3D analysis along with a tomographic one, showing the increase in sensitivity of the former.

In our analysis we did not consider spurious contributions to the pure lensing signal coming from systematics such as the intrinsic alignments of source galaxies \citep{Joachimi2010, Mandelbaum2017}. These are expected to dominate the error budget for future cosmic shear surveys and need therefore to be carefully accounted for. These contributions are also expected to influence mostly the lensing signal on small, non-linear scales. The scales we considered were also chosen with the purpose of avoiding the regime of domination of these effects, which we considered neither in our fully 3D approach nor in the tomographic one, so that the comparison could remain fair. However, it has been shown that this kind of systematics can be carefully accounted for in 3D analyses \citep{Merkel2013} and we plan to investigate their impact in future work, together with cross-correlations with other probes which we envisage as one of the most powerful tools to test gravity on cosmological scales. 
\section*{Acknowledgements}
\addcontentsline{toc}{section}{Acknowledgements}

We thank Thomas Kitching, Peter Taylor and Britta Zieser for useful discussions and comments. 

ASM and RR acknowledge financial support from the graduate college \textit{Astrophysics of cosmological probes of gravity} by Landesgraduiertenakademie Baden-W\"urttemberg. MZ is supported by the Marie Sklodowska-Curie Global Fellowship Project NLO-CO.

\bibliographystyle{mnras}
\bibliography{references} 

\appendix
\section{Varying $\hat{\alpha}_K$ and $\hat{\alpha}_T$}

\begin{figure*}
	\begin{center}
		\includegraphics[width = \textwidth]{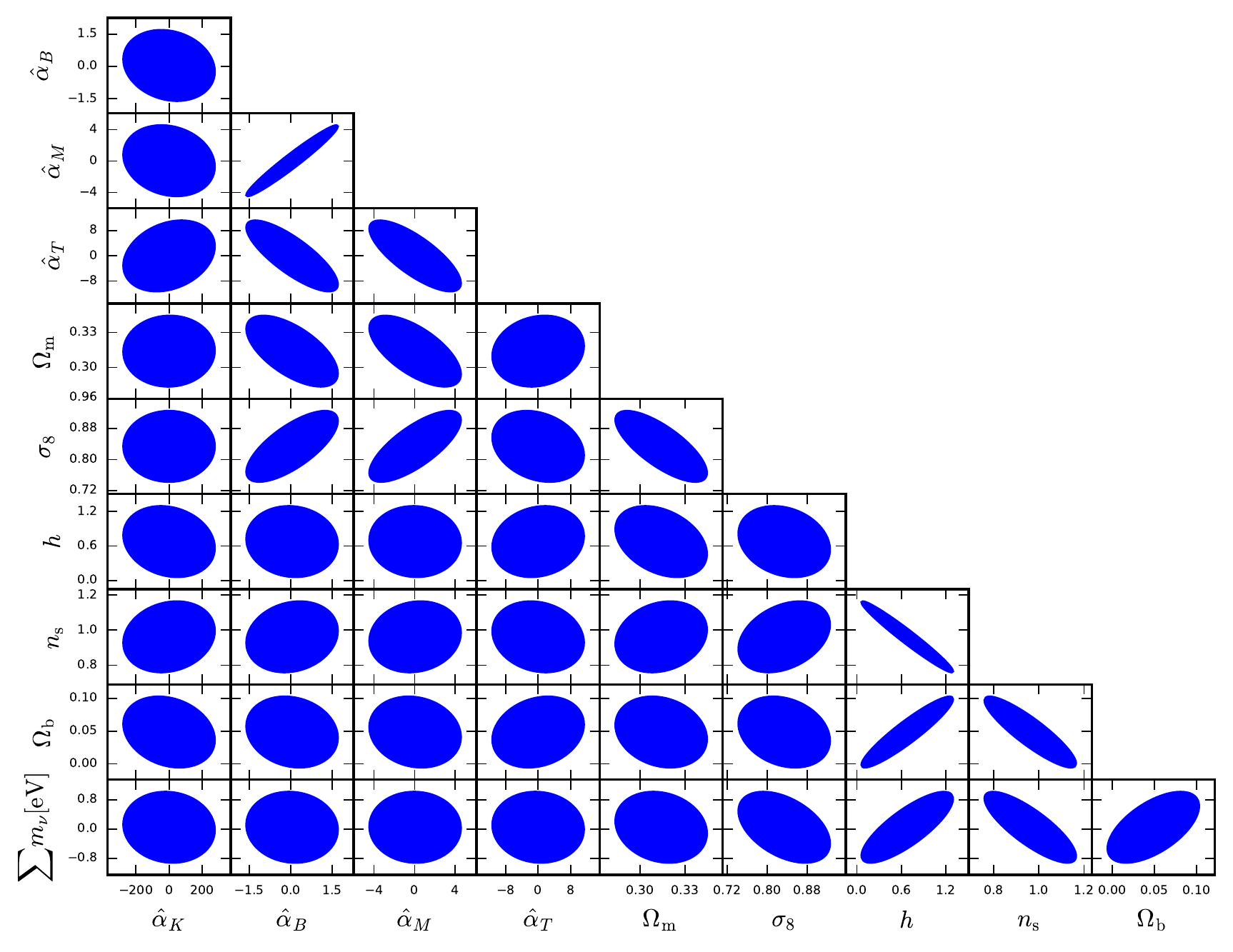}
		\caption{Fiducial values from \autoref{tab:parameters_ak001_tomo3Dlin} and survey specifications are given in \autoref{tab:specifications}. We notice that $\hat{\alpha}_K$ is unconstrained.}
		\label{fig:varying_alphaK_alphaT}
	\end{center}
\end{figure*}

We show in \autoref{fig:varying_alphaK_alphaT} the contour plots that we obtain if we also vary $\hat{\alpha}_K$ and $\hat{\alpha}_T$. We notice in particular that, as expected, $\hat{\alpha}_K$ is unconstrained, therefore we decide to fix it at its fiducial value.

\section{Influence of $k_V$}\label{app:kv}
\begin{figure}
\begin{center}
\includegraphics[width = .45\textwidth]{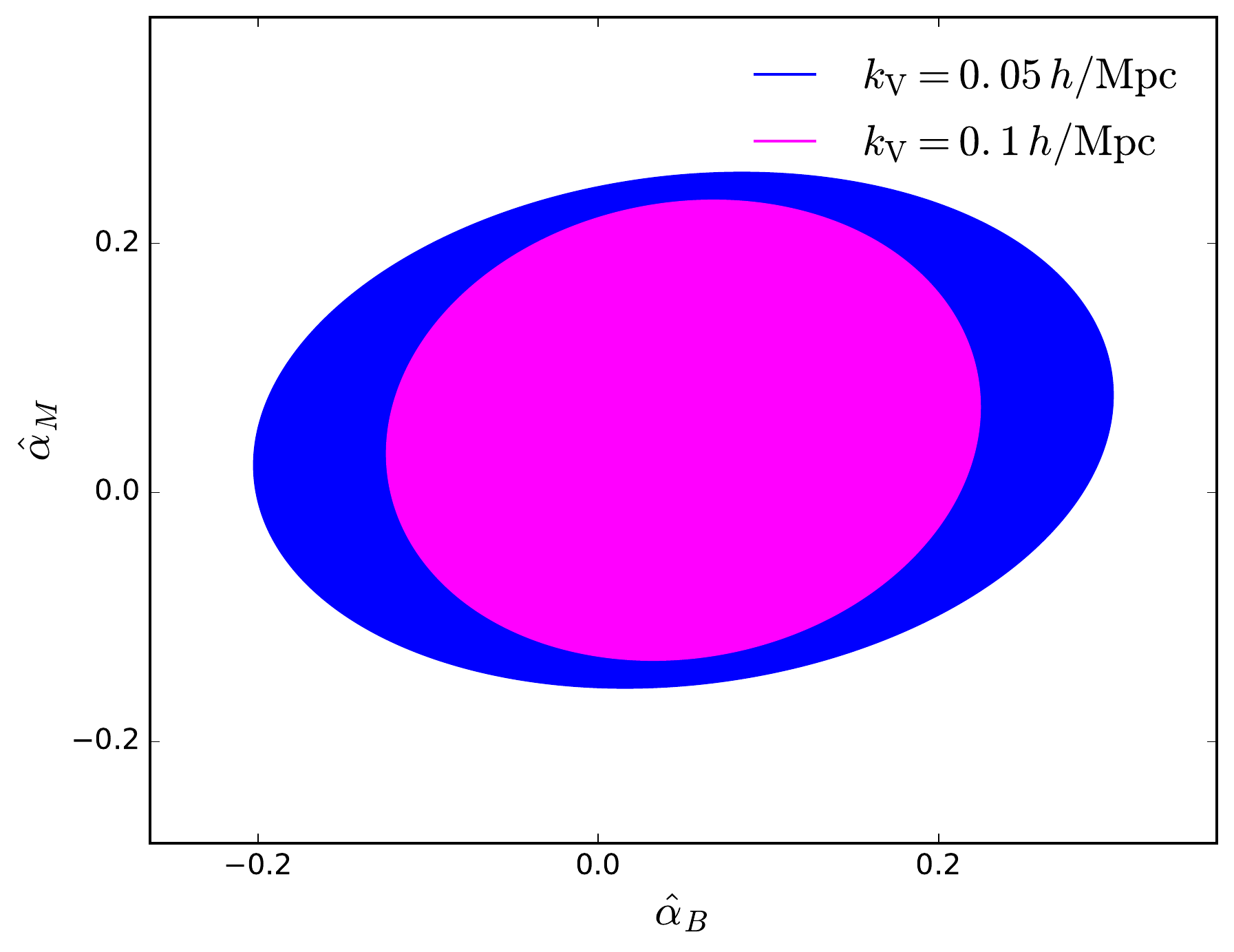}
\caption{Constraints on $\hat{\alpha}_B$ and $\hat{\alpha}_M$ for two different choices of the screening length $k_V$.}
\label{fig:screening}
\end{center}
\end{figure}

In \autoref{fig:screening} we show the influence of the screening length on the constraining power. Clearly, if $k_V$ becomes smaller, GR is retained at larger scales already, thus decreasing the sensitivity on the modified gravity parameters. For a more complete discussion we refer the reader to \citet{Alonso2016}.

\bsp
\label{lastpage}

\end{document}